\documentclass[journal,twoside]{IEEEtran}
\usepackage{amsmath}%\documentclass[12pt, draftclsnofoot, onecolumn]{IEEEtran}
\usepackage{amsfonts}

\usepackage{multirow}
\usepackage{footnote}
\usepackage{epsfig}
\usepackage{amssymb}
\usepackage{amsmath}
\usepackage{cite}
\usepackage{color,soul}
\usepackage{subfigure}
\usepackage{multicol}
\usepackage{color}
\usepackage{rotating}
\usepackage{graphicx}
\usepackage{tabularx}
\usepackage{array}
\usepackage{color,soul}
\usepackage{graphicx,dblfloatfix}
\usepackage{blindtext}
\usepackage{ragged2e}

\usepackage{algorithmic}
\usepackage[Algorithm]{algorithm}

\begin{document}
    \title{Robust Physical Layer Security for Power Domain Non-orthogonal
    	Multiple Access-Based HetNets and HUDNs: SIC Avoidance at Eavesdroppers}
    \author{\IEEEauthorblockN{Moslem Forouzesh, Paeiz Azmi, \textit{Senior Member, IEEE,} Nader Mokari, \textit{Member, IEEE,} and Kai Kit Wong, \textit{Fellow, IEEE}}  \textsuperscript{}\thanks{\noindent\textsuperscript{} Moslem Forouzesh is with the Department of Electrical and
    		Computer Engineering, Tarbiat Modares University, Tehran, Iran
    		(e-mail: m.Forouzesh@modares.ac.ir).
    		
    		Paeiz Azmi is with the Department of ECE, Tarbiat Modares University,
    		Tehran, Iran (e-mail: pazmi@modares.ac.ir).
    		
    		Nader Mokari is with the Department of ECE, Tarbiat Modares University,
    		Tehran, Iran (e-mail: nader.mokari@modares.ac.ir).
    		
    		Kat-Kit Wong is with the Department of Electronic and Electrical Engineering, University College London, WC1E 7JE, United
    		Kingdom (e-mail: kai-kit.wong@ucl.ac.uk).}}
    \maketitle
%    in contrast to conventional modes, that destinatin is halfe duplex

    \begin{abstract}
       In this paper, we investigate the physical layer security in downlink of  Power Domain Non-Orthogonal
        Multiple Access (PD-NOMA)-based  heterogeneous cellular network (HetNet) in presence of multiple eavesdroppers.  Our aim  is to maximize the sum secrecy  rate of the network. To this end, we formulate joint subcarrier and power allocation optimization problems to increase sum secrecy  rate. Moreover, we propose a novel scheme at which the eavesdroppers are prevented  from doing  Successive Interference Cancellation (SIC),  while  legitimate users are able to do it. 
        In practical systems, availability of eavesdroppers' Channel State Information (CSI) is impractical, hence we consider two scenarios: 1) Perfect CSI of the eavesdroppers, 2) imperfect CSI of the eavesdroppers.
  Since the  proposed optimization  problems are non-convex, we adopt the well-known iterative algorithm called Alternative Search Method (ASM). In this algorithm, the optimization problems are converted to two subproblems, power allocation and subcarrier allocation. We  solve the power allocation problem  by the Successive Convex Approximation approach and solve the subcarrier allocation subproblem,   by exploiting the Mesh Adaptive Direct Search algorithm (MADS).
        Moreover,  in order to study the optimality gap of the proposed solution method, we  apply the monotonic optimization method. 
	Moreover, we evaluate the proposed scheme for secure massive connectivity
		 in 5G networks.
        Numerical results highlight that the proposed scheme significantly improves the sum secrecy  rate compared with the conventional case at which the eavesdroppers are able to apply SIC. 
     \end{abstract}
    \section{Introduction}\label{introduction}
    \subsection{State of the art and motivation}
    \IEEEPARstart{T }{he}
\textcolor{black}{increasing demand of  high data rates and multimedia applications and scarcity of radio resources encourage operators, research centers, and vendors  to devise new methods and products for providing high data rate services for the next-generation 5G network.  International Telecommunication Union (ITU)  has categorized 5G services into three categories: 1) Ultra-reliable and low latency communication (URLLC), 2) enhanced mobile broadband (eMBB), and 3) massive machine-type
communication (mMTC) \cite{ITU}. In mMTC, massive number of machine-type devices are connected simultaneously.  Services like sensing, monitoring, and tagging which are in this category have two main challenges: 1) Scarcity of radio resources which make the deployment  massive connections very difficult and 2) broadcast nature of wireless channels which make the massive connections insecure. 
Massive connectivity is one the main features of future wireless cellular networks which is suitable for IoT \cite{J.Zhu} and machine-type communications (MTC) services.
In massive device connectivity scenarios, a cellular BS  connects to  large number of devices (in order of $10^4$ to $10^6$ per $\text{Km}^2$). To achieve high throughput and spectrum efficiency, Heterogeneous Ultra Dense Networks (HUDNs) is a promising solution at which the  number of BSs per $\text{Km}^2$ is very large (about 40-50) \cite{Z.Qin}.
In order to overcome the scarcity of radio resources in this category,  a new multiple access method called power domain Non-Orthogonal Multiple Access  (PD-NOMA) can be adopted at which users are serviced within a given resource slot (e.g., time/frequency) at different 
levels of transmit power \cite{Y. Saito_i}, \cite{M. Al-Imari}. In this method, users can remove  signals intended for other users which have the worse channel conditions, by employing successive
interference cancellation (SIC) \cite{T. Cover, T. M. Cover, J. G. Andrews}. 
It is necessary to mention that SIC concept was first proposed by Cover in 1972 \cite{T. M. Cover} which is very useful technique, because it imposes lower complexity than joint
decoding techniques \cite{J. G. Andrews}. }
It is worth noting that the PD-NOMA technique has attracted significant attentions in  both academia and industry, \cite{Z. Ding_i, moltafet, pout, mmw}.
It is necessary to mention that it has been confirmed in theory domain \cite{L.Dai} and system-level simulations \cite{Y. Saito} which  PD-NOMA surpasses orthogonal frequency-division multiple access (OFDMA) in different points of view such as device connections and  spectrum efficiency.
 Based on these benefits, PD-NOMA is very appropriate to be employed for meeting the 5G requirements such as  massive connectivity \cite{Z.Ding} which is  very vital for mMTC and Internet of Things (IoT).  
Besides, establishment of security in these networks is a dilemma, because wireless transmission has broadcast nature. Therefore, private information that is exchanged between transmitter and receiver is vulnerable of eavesdropping. During the past years, physical layer security as a promising idea, 
has been widely investigated since Wyner’s presented his work  in the security domain\cite{A. D. Wyner}. Furthermore, as IoT is employed in wide domains such as commercial,  military, and governmental application,
	 security  plays an important role in  IoT applications \cite{Granjal}. ِDue to constraints of  energy consumption and limited hard-ware in IoT devices, it is very vulnerable with respect to eavesdropping. Physical layer security owing to low computational complexity attracts a lot of attentions and is becoming a suitable solution  for secure communications in IoT \cite{N.Yang}.  
    \subsection{Related works}
    In recent years, PD-NOMA has been studied from various perspectives, such as investigating optimal and fair energy efficient resource allocation in downlink of  	Heterogeneous Network (HetNet) for energy harvesting \cite{moltafet},  
   outage performance analysis \cite{pout}, and  beamforming based 5G millimeter-wave communications  \cite{mmw}. Moreover, power allocation  ensuring fairness
  among users for 
  instantaneous and average CSI, is investigated in \cite{S. Timotheou}. In 
%\cite{J. Choi},
\cite{Z. Ding.R. Schober}, PD-NOMA based multiple antennas
technology is proposed  to improve throughput. Multiple input single-output (MISO) in  PD-NOMA-based network is investigated in \cite{J. Choi}. \textcolor{black}{Additionally, in \cite{Z.Qin}, the authors study usage
of PD-NOMA in HUDNs to backup massive connectivity in 5G networks. Cost of active user detection and channel estimation in massive connectivity by employing massive MIMO is evaluated in \cite{L.Liu}. The authors in \cite{P.Soldati}, propose an inter-cell interference coordination mechanism  in dense small cell networks. Millimeter-Wave PD-NOMA in  machine-to-machine (M2M) communications for IoT networks is proposed in \cite{LvT}. The authors in \cite{Zhai} study dynamic user scheduling
and power allocation problem for massive IoT devices based PD-NOMA. }

 Recently, physical layer security in single-input single-output (SISO) systems based on the PD-NOMA technology is investigated in \cite{Y.Z}, at which its objective is maximizing the sum secrecy  rate.
 Physical layer security for PD-NOMA-based cognitive radio networks is studied in \cite{Xiang}. The authors' aim is to derive exact and asymptotic expressions of the outage secrecy rate. 
 Comparison secrecy unicasting rate between PD-NOMA and OMA is investigated in  \cite{Ding.Z}. The authors study the achievable secrecy unicasting rate of 
 OMA and PD-NOMA.
  Moreover, in \cite{Liu. Yuanwei}, the authors study  physical layer security of PD-NOMA in  large-scale networks with employing stochastic geometry, in which a new exact expression of the outage secrecy rate  is derived for single-antenna and multiple-antenna cases. In the literature, it is ssumed that eavesdroppers know the channel ordering and are able to perform SIC. 
It is worth noting that eavesdroppers by doing SIC are able to decrease the sum secrecy  rate. Hence, to tackle this issue, we propose  a novel scheme such that,
we do not allow  the eavesdroppers to be able to perform SIC, even if they know the channel ordering. Moreover, based on aforementioned discussions, there are few works in the scope of integration of PD-NOMA and physical layer security in wireless networks.

\subsection{Contribution}

In the following, we summarize the main contributions of this paper as:

\begin{itemize}
	
	\item
	We consider physical layer security for the downlink of 
	PD-NOMA-based HetNets and HUDNs consisting of one Macro Base Station (MBS), multiple  Small Base Stations (SBS), multiple MBS,  SBS users, and multiple eavesdroppers.
	
	\item
	We focus on this aspect of PD-NOMA, ``how to avoid eavesdroppers from doing SIC and when users are able to perform SIC?". 
	From the information theory point of view, user $A$ can perform SIC whenever its 
	received Signal-to-Interference-plus-Noise Ratio (SINR) at user $B$ is more than  user $B$'s received SINR at its own signal,  \cite{Z. Ding_interf}, \cite{D. Tse and P. Viswanath}. We propose  a  novel optimization problem such that,
	we do not allow   eavesdroppers to be able to perform SIC, even if they know the channel ordering. In this regard, we formulate an optimization problem at which we introduce a new constraint called SIC avoidance at  eavesdropper condition and the main aim is to maximize the sum secrecy  rate over transmit power and subcarrier allocation variables. 

	\item
	For solving  the proposed optimization problems, we adopt the well-known  iterative algorithm called Alternative Search Method (ASM) \cite{K. Son}. In this method, the optimization problems are converted to two subproblems which one of them has binary  and another has  continuous optimization variables, in  other words, power and subcarriers are allocated separately. In each iteration of this method, power allocation problem
	is  non-convex  and subcarrier allocation is   Non-Linear Programing (NLP). We  solve the power allocation problem by the Successive Convex Approximation
	(SCA) approach. To this end, we use Difference of two Concave functions
	(DC) approximation, to transform the non-convex problem into a
	canonical form of convex optimization \cite{P. Parida}. Also we solve subcarrier allocation problem, by exploiting the Mesh Adaptive Direct Search algorithm (MADS), to this end, we employ  Nonlinear Optimization solver with  MADS  (NOMAD), \cite{http}.
	\item 
	In practical systems, availability of all eavesdroppers' channel conditions at legitimate transmitters are impractical, hence, we investigate two scenarios: 1) Perfect Channel State Information (CSI) of eavesdropper, 2) imperfect CSI of eavesdropper, where BSs do not have perfect CSI of eavesdroppers.
	\item
	 We also apply the monotonic optimization method to study the optimality gap of the proposed solution method \cite{48}. For this purpose, we convert the optimization problems to the canonical form of monotonic optimization problem and finally, by exploiting the polyblock algorithm, we solve the monotonic optimization problem, globally.
	 
	 \item
	\textcolor{black}{We also evaluate  the proposed scheme for secure massive connectivity in 5G ultra dense networks.
	 Without loss of generality, for changing our scenario from HetNet to HUDN, we need to extend the dimension of system model. According to \cite{Z.Qin} and \cite{P.Soldati}, in order to tackle high dimension complexity of resource allocation in HUDNs,  it is assumed that the transmit power is uniformly allocated to users. Moreover, we show the performance of uniform power allocation is close to the performance of our proposed solution.}
	
\end{itemize}
\subsection{Organization}
The remainder of this paper is organized as follows. In Section \ref{systemmodel}, we present system and signal model, respectively, and explain our novel idea. Section \ref{Problem Formulation} provides the detailed problem formulation at two scenarios: 1) Perfect CSI, 2) imperfect CSI. In Section \ref{SOLUTIONS OF THE OPTIMIZATION PROBLEM}, the proposed solution is expressed. 
The proposed scheme for ultra dense network is evaluated in Section \ref{MCS}.
Our proposed optimal solution is provided in Section \ref{Optimal solution}. Performance evaluation of the proposed resource allocation approach is discussed in Section \ref{SIMULATION RESULTS}, before ending, the paper is
concluded in Section \ref{Conclusion}.
%%%%%%%%%%%%%%%%%%%%%%%%%%%%%
\begin{figure}[b!]
	\centering
	\includegraphics[width=3.5in,height=2.4in]{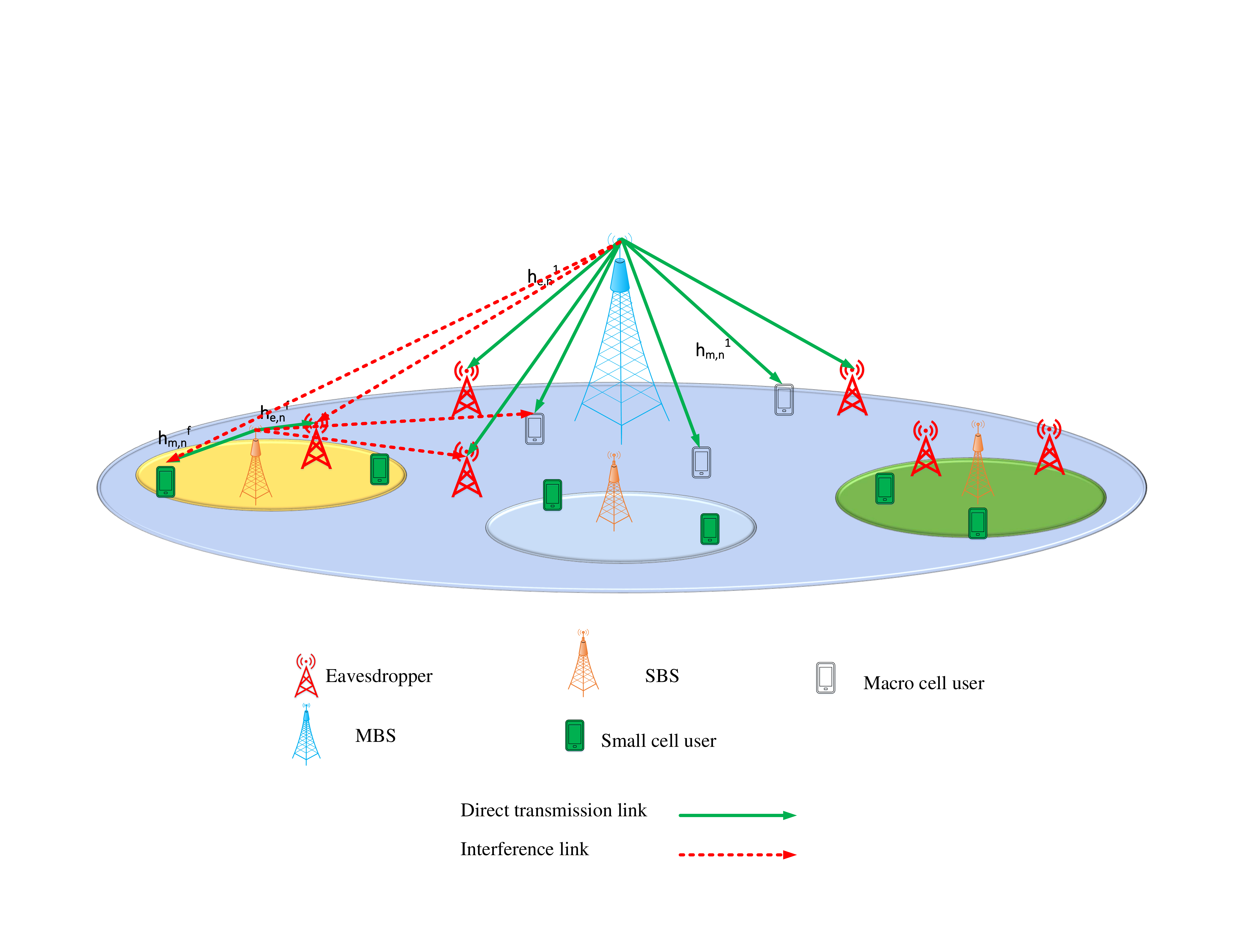}% \vspace{3cm}
	\caption{Secure transmission in downlink of PD-NOMA based HetNet.}
	\label{System_M}
\end{figure}

\begin{center}
	\begin{table}[h]
		
		\caption{List of The Main Variables}
		\begin{tabular}{ | c || p{6cm} |}
			\hline
			variables & \hspace{2cm} definition\\
			
			\hline \hline
			
			\hline
			${h_{m,n}^f}$ & {\small  Channel
				coefficient from BS  $f$ to the $m^{th}$ user on subcarrier  $n$}    \\
			\hline
			${h_{e,n}^f}$ & {\small Channel
				coefficient from BS  $f$ to the $e^{th}$ eavesdropper on subcarrier  $n$}    \\
			\hline
			${p_{m,n}^f}$ & {\small Transmit power  of BS $f$ to user  $m$ on subcarrier $n$} \\
			\hline
			
			${\rho_{m,n}^f}$ & {\small Subcarrier allocation to user  $m$ on subcarrier $n$
				in BS $f$}  \\
			\hline
			
			${\mathcal{F}}$ & {\small  Set of BSs, $ \mathcal{F} =\left\{ {1,2,...F} \right\}$} \\
			\hline
				${F}$ & {\small  Total  number of BSs}\\
			\hline
			${\mathcal{N}}$ & {\small Set of total subcarriers,
				$\mathcal{N}= \left\{ {1,2,...N} \right\}$} \\
			\hline
			
			${N}$ & {\small  Total  number of subcarriers}\\
			
			\hline
			
			${\mathcal{M}_f}$ & {\small Set of all users in BS $f$,  $\mathcal{M}_f= \left\{ {1,2,...M_f} \right\}$ }\\
			\hline
			
			${M}$ & {\small  Total number of users,  $M=\sum_{f \in \mathcal{F}}^{}{M}_{f} $ }\\
			\hline
			
			${\mathcal{E}}$ & {\small Set of all eavesdroppers, $\mathcal{E}= \left\{ {1,2,...E} \right\}$} \\
			\hline
			
			${E}$ & {\small  Total number of eavesdroppers}. \\
			\hline
		\end{tabular}
		\label{variable table}
	\end{table}
\end{center}
\section{System and Signal Model}\label{systemmodel}
\subsection{System model}

In this paper, we focus on secure communication in the downlink of PD-NOMA based HetNet. As illustrated in Fig. \ref{System_M},  our system model consists of one MBS, multiple SBSs, multiple eavesdroppers, multiple MBS,  and SBS users. In this system model, we assume that all nodes are equipped with single antenna.
For clarity, the main underutilized variables in this paper are listed in Table \ref{variable table}. Note that $f=1$ refers  to MBS. When  BS $f$ allocates subcarrier $n$ to user $m$, the binary variable $\rho_{m,n}^f \in \left\{ {0,1} \right\}$ is equal to one, i.e., $\rho_{m,n}^f=1$, and  otherwise, $\rho_{m,n}^f=0$.  $h_{m,n}^f= d_{m,f}^{-\alpha}{\tilde h_{m,n}^f}$ is the channel coefficient between BS $f$ and user $m$ on subcarrier $n$, where ${\tilde h_{m,n}^f}$ indicates the Rayleigh
fading, $\alpha$  and $d_{m,f}$ are the path loss exponent and the distance between user $m$ and BS $f$, respectively. َ
Moreover, $h_{e,n}^f= d_{e,f}^{-\alpha}{\tilde h_{e,n}^f}$.

\subsection{Signal model}
Employing the PD-NOMA technique,  a linear combination of $M_f$ signals is diffused by BS $f$ to its users, \cite{Z. Ding_interf}. In other words, BS $f$ transmits $\sum\limits_{j = 1}^{{M_f}} {\rho _{j,n}^f\sqrt {p_{j,n}^f} } s_{j,n}^f$ on  subcarrier $n$, where $s_{j,n}^f$  denotes  the transmitted symbol of the $j^{th}$ user on the $n^{th}$ subcarrier by BS $f$. Without loss of generality, it is assumed ${\rm E}\left\{ {{{\left| {s_{m,n}^f} \right|}^2}} \right\} = 1, \forall  m\in \mathcal{M}_f,  f \in \mathcal{F},  n \in \mathcal{N}$,  where $\mathcal{M}_f$, $\mathcal{N}$, and $\mathcal{F}$ are denoted set of all users in BS, set of total subcarriers, and set of BSs, respectively. Moreover,  ${\rm E}\left\{ {{ x }} \right\}$ is the expectation of $x$. The received signals   on the  $m^{th}$ user  and the $e^{th}$ adversary, that are located in the coverage region of BS $f$,  on  the $n^{th}$ subcarrier are expressed as
\begin{align}
&	y_{m,n}^f = h_{m,n}^f\sum\limits_{i \in \mathcal{M}_f}^{{}} {\rho _{i,n}^f\sqrt {p_{i,n}^f} } s_{i,n}^f + Z_{m,n}^f+\\& \sum\limits_{f^\prime \in \mathcal{F}/f}^{{}}{h_{m,n}^{f'}}\sum\limits_{i \in \mathcal{M}_{f^\prime}}^{{}}{\rho _{i,n}^{f'}\sqrt {p_{i,n}^{f'}} s_{i,n}^{f'}}\nonumber
\end{align}
and
\begin{align}
&	y_{e,n}^{f} = h_{e,n}^f\sum\limits_{i \in \mathcal{M}_f}^{{}} {\rho _{i,n}^f\sqrt {p_{i,n}^f} } s_{i,n}^f + Z_{e,n}^f+\\&\sum\limits_{f^\prime \in \mathcal{F}/f}^{{}}{h_{e,n}^{f'}}\sum\limits_{i \in \mathcal{M}_{f^\prime}}^{{}}{\rho _{i,n}^{f'}\sqrt {p_{i,n}^{f'}} s_{i,n}^{f'}} , \nonumber
\end{align}
respectively, 
where $Z_{m,n}^f\sim \mathcal{C}\mathcal{N}\left( {0,\sigma^2} \right)$ and $Z_{e,n}^f\sim \mathcal{C}\mathcal{N}\left( {0,\sigma^2} \right)$ 
are the complex Additive White Gaussian Noise (AWGN) with zero-mean and variance $\sigma^2$, on subcarrier $n$, over BS $f$, at user $m$ and eavesdropper $e$, respectively.

\subsection{Achievable Rates at the Legitimate Users and the Eavesdroppers}
In order to decode signals, in  the PD-NOMA-based system,  users apply SIC  \cite{Y. Saito}, \cite{L. Lei}. In these systems, user $m$,  firstly detects the $i^{th}$ user's message when ${{\left| {h_{m,n}^f} \right|}^2} > {{\left| {h_{i,n}^f} \right|}^2}$, then removes  the detected message from  the received signal, in a consecutive way. Note that  the $i^{th}$ user's message   for user $m$ behaves as noise if ${{{\left| {h_{m,n}^f} \right|}^2} \le {{\left| {h_{i,n}^f} \right|}^2}}$. When user $m$ applies  SIC, its SINR in BS $f$ on subcarrier $n$  can be expressed as:

\begin{align}\label{SINR_m}
	\gamma _{m,n}^f = \frac{{p_{m,n}^f{{\left| {h_{m,n}^f} \right|}^2}}}{{{{\left| {h_{m,n}^f} \right|}^2}\sum\limits_{\scriptstyle{\left| {h_{m,n}^f} \right|^2} \le {\left| {h_{i,n}^f} \right|^2}\hfill\atop
				\scriptstyle i \in {\mathcal{M}_f}/\left\{ m \right\}\hfill}^{} {p_{i,n}^f\rho _{i,n}^f}  +I_{m,n}^f+ {\sigma^2}}},
\end{align}
where $I_{m,n}^f = \sum\limits_{{f^\prime } \in {\cal F}/f}^{} {{\left| {{h_{m,n}^{f'}}} \right|}^2} \sum\limits_{i \in {{\cal M}_{{f^\prime }}}}^{} {\rho _{i,n}^{f'}p_{i,n}^{f'}} $
and its achievability rate is given by:
\begin{align}
	r_{m,n}^f = \log (1 + \gamma _{m,n}^f).
\end{align}
In the PD-NOMA-based system, users are able to perform SIC,  if the following conditions hold \cite{Sun}, \cite{B. R. Marks}:

\begin{itemize}
	\item  SIC can be applied by user $m$ if user $i$'s received SINR for its signal is less than or equal to user $m$'s received SINR for  user $i$'s signal \cite{Z. Ding_interf,D. Tse and P. Viswanath}.
	\item
	$\vert h_{i,n}^{f}\vert^2 \leq \vert h _{m,n}^{f}\vert^2, \forall i, m\in \mathcal{M}_f,  f \in \mathcal{F},  n \in \mathcal{N}, i \neq m.  $
\end{itemize}
In other words, user $m$ can successfully decode and remove the $i^{th}$ user's signal on subcarrier $n$ in BS $f$,  whenever the following inequality is  satisfied, \cite{T. Wang}: 
\begin{align}\label{SIC_Condition1}
	&\gamma _{m,n}^f\left( i \right) \ge \gamma _{i,n}^f\left( i \right) \forall i, m\in \mathcal{M}_f,  f \in \mathcal{F},  n \in \mathcal{N}, \\& \vert h_{i,n}^{f}\vert^2 \leq \vert h _{m,n}^{f}\vert^2, i \neq m, \nonumber 
\end{align}
where $\gamma _{m,n}^f\left( i \right) $ is user $m$'s SINR for user $i$'s signal
and $\gamma _{i,n}^f\left( i \right)$ is  user $i$'s SINR for its own signal. Accordingly, \eqref{SIC_Condition1} can be written as follows:
\begin{align}\label{SIC_Condition2}
	&{\log _2}\left(1+ {\frac{{p_{i,n}^f{{\left| {h_{m,n}^f} \right|}^2}}}{{{{\left| {h_{m,n}^f} \right|}^2}\sum\limits_{\scriptstyle{\left| {h_{m,n}^f} \right|^2} \le {\left| {h_{l,n}^f} \right|^2}\hfill\atop
					\scriptstyle l \in {\mathcal{M}_f}/\left\{ i \right\}\hfill}^{} {p_{l,n}^f\rho _{l,n}^f}  +I_{m,n}^f+ {\sigma^2}}}} \right) \ge \nonumber\\& {\log _2}\left(1+ {\frac{{p_{i,n}^f{{\left| {h_{i,n}^f} \right|}^2}}}{{{{\left| {h_{i,n}^f} \right|}^2}\sum\limits_{{{\left| {h_{i,n}^f} \right|}^2} \le {{\left| {h_{l,n}^f} \right|}^2}\hfill\atop
					\scriptstyle l \in {\mathcal{M}_f}/\left\{ i \right\}\hfill}^{} {p_{l,n}^f\rho _{l,n}^f}  + I_{i,n}^f+{\sigma^2}}}} \right),
\end{align}
for simplicity,  we can rewrite  \eqref{SIC_Condition2} to the following inequality:
\begin{equation}\label{SIC_linear-u}
	\begin{split}
		& Q_{m,i,n}^f({\boldsymbol{\rho }},{\bf{p}}) \buildrel \Delta \over =  - |h_{m,n}^f{|^2}{\sigma^2} + |h_{i,n}^f{|^2}{\sigma^2}+|h_{i,n}^f{|^2}I_{m,n}^f-\\&|h_{m,n}^f{|^2}I_{i,n}^f - {\left| {h_{m,n}^f} \right|^2}{\left| {h_{i,n}^f} \right|^2}\sum\limits_{{{\left| {h_{i,n}^f} \right|}^2} \le {{\left| {h_{l,n}^f} \right|}^2}\hfill\atop
			\scriptstyle l \in {\mathcal{M}_f}/\left\{ i \right\}\hfill}^{} {p_{l,n}^f\rho _{l,n}^f}  +\\& {\left| {h_{i,n}^f} \right|^2}{\left| {h_{m,n}^f} \right|^2}\sum\limits_{{{\left| {h_{m,n}^f} \right|}^2} \le {{\left| {h_{l,n}^f} \right|}^2}\hfill\atop
			\scriptstyle l \in {\mathcal{M}_f}/\left\{i  \right\}\hfill}^{} {p_{l,n}^f\rho _{l,n}^f}  \le 0.
	\end{split}
\end{equation}

We propose a novel resource allocation algorithm in which  eavesdropper is not able to employ SIC to increase its own achievable rate.  
%Our basic idea is to allocate resources
%such that the eavesdropper $e$'s received SINR for user $m$'s signal is less than  user $m$'s received SINR for its own signal. 
In this case, eavesdropper $e$ can not apply SIC, hence,  all users' messages are treated as interference in the $e^{th}$ eavesdropper. 
Therefore, SINR of the eavesdropper $e$ in BS $f$ on subcarrier $n$ can be obtained as:
\begin{align}\label{SINR_e}
	\gamma _{e,n}^{f,m} = \frac{{p_{m,n}^f{{\left| {h_{e,n}^f} \right|}^2}}}{{{{\left| {h_{e,n}^f} \right|}^2}\sum\limits_{
				\scriptstyle{i\in \mathcal{M}_f/\left\{ m \right\}}\hfill}^{} {p_{i,n}^f\rho _{i,n}^f}  +I_{e,n}^f+ {\sigma^2}}}.
\end{align}
where $I_{e,n}^f = \sum\limits_{{f^\prime } \in {\cal F}/f}^{} {{\left| {{h_{e,n}^{f'}}} \right|}^2} \sum\limits_{i \in {{\cal M}_{{f^\prime }}}}^{} {\rho _{i,n}^{f'}p_{i,n}^{f'}} $
and its achievable rate is given by:
\begin{align}
	r_{e,n}^{f,m} = \log (1 + \gamma _{e,n}^{f,m}),
\end{align}
For SIC avoidance  at the eavesdroppers, the following inequality must be satisfied:
\begin{align}\label{SIC_Condition_e1}
	&\gamma _{e,n}^{f,m}\left( i \right) \le \gamma _{i,n}^f\left( i \right), \forall i, m\in \mathcal{M}_f,  e\in \mathcal{E}, f \in \mathcal{F},  n \in \mathcal{N}\\& \vert h_{i,n}^{f}\vert^2 \leq \vert h _{e,n}^{f}\vert^2, i \neq m, \nonumber
\end{align}
in other words:
\begin{align}\label{SIC_Condition_e2}
	&{\log _2}\left(1+ {\frac{{p_{i,n}^f{{\left| {h_{e,n}^f} \right|}^2}}}{{{{\left| {h_{e,n}^f} \right|}^2}\sum\limits_{l \in {{\cal M}_f}/\left\{ i \right\}}^{} {p_{l,n}^f\rho _{l,n}^f}  + I_{e,n}^f+{\sigma^2}}}} \right) \le \nonumber\\&{\log _2}\left(1+ {\frac{{p_{i,n}^f{{\left| {h_{i,n}^f} \right|}^2}}}{{{{\left| {h_{i,n}^f} \right|}^2}\sum\limits_{{{\left| {h_{i,n}^f} \right|}^2} \le {{\left| {h_{l,n}^f} \right|}^2}\hfill\atop
					\scriptstyle l \in {\mathcal{M}_f}/\left\{ i \right\}\hfill}^{} {p_{l,n}^f\rho _{l,n}^f}  +I_{i,n}^f+ {\sigma^2}}}} \right),
\end{align}
  by some mathematical manipulation, \eqref{SIC_Condition_e2} is equivalent to the following inequality:
\begin{align}
	&\Psi _{m,i,n,e}^f({\boldsymbol{\rho }},{\bf{p}}) =
	- |h_{e,n}^f{|^2}{\sigma^2} + |h_{i,n}^f{|^2}{\sigma^2}-|h_{e,n}^f{|^2}I_{i,n}^f+ \nonumber \\&|h_{i,n}^f{|^2}I_{e,n}^f-
	{\left| {h_{e,n}^f} \right|^2}{\left| {h_{i,n}^f} \right|^2}\sum\limits_{{{\left| {h_{i,n}^f} \right|}^2} \le {{\left| {h_{l,n}^f} \right|}^2}\hfill\atop
		\scriptstyle l \in {\mathcal{M}_f}/\left\{ i \right\}\hfill}^{} {p_{l,n}^f\rho _{l,n}^f}  + \\&
	{\left| {h_{i,n}^f} \right|^2}{\left| {h_{e,n}^f} \right|^2}\sum\limits_{l \in {\mathcal{M}_f}/\left\{ i \right\}}^{} {p_{l,n}^f\rho _{l,n}^f}  \ge 0.\nonumber
\end{align}
In the following, we assume that the eavesdroppers are non-colluding, hence we have
\begin{align}
	r_{e_{\max} ,n}^{f,m} = {\max _{e \in \mathcal{E} }}\left\{ {\log (1 + \gamma _{e,n}^{f,m})} \right\},
\end{align}
therefore, the secrecy rare at the $m^{th}$ user served by BS $f$ on subcarrier $n$ can be obtained as follows:
\begin{align}
	R_{m,n}^{\sec f} = {\left[ {r_{m,n}^f - r_{e_{\max} ,n}^{f,m}} \right]^ + },
\end{align}
where $\left[ \Psi  \right]^ + = \max \left\{ {\Psi ,0} \right\}$.
\section{Problem Formulation}\label{Problem Formulation}
In this section, we propose a new policy for resource allocation to maximize the sum secrecy  rate. It should be noted, in practical systems, having knowledge of all eavesdroppers' CSI  is impractical, hence we investigate two scenarios: 1) Perfect CSI of the eavesdroppers, 2) imperfect CSI of the eavesdroppers, where BSs do not have perfect CSI of eavesdroppers. We investigate these two scenarios in two Subsections \ref{perfect CSI} and \ref{Imperfect CSI}, respectively.

\subsection{Perfect CSI Scenario}\label{perfect CSI}
In this subsection, we assume the CSI of eavesdroppers are available in the BSs, which is a common assumption in the physical layer security literature \cite{I_canselation1}\footnote{It is assumed the eavesdroppers are users of network, which are not legitimate for accessing some information.}. To this end, we propose a policy for resource allocation to maximize the sum secrecy  rate. In this policy, unlike the users, the eavesdroppers cannot apply SIC. We formulate the considered optimization problem of sum secrecy  rate maximization  via the worst-case robust approach as follows:

\begin{subequations}\label{Opti_prob}
	\begin{align}
		&\max_{\mathbf{P},\boldsymbol{\rho}}\; \sum_{\forall f\in\mathcal{F}}\sum_{\forall m\in \mathcal{M}_f}\sum_{\forall n\in \mathcal{N}}\rho^f_{m,n} \left\{ {r_{m,n}^f - r_{e_{\max} ,n}^{f,m}} \right\},\\& \label{Opti_probb}
		\hspace{-.5cm}\text{s.t.}:\hspace{.18cm} C_1:\hspace{.002cm}
		\sum_{m\in \mathcal{M}_f}\sum_{n\in \mathcal{N}}\rho^f_{m,n}p^f_{m,n}\le p^f_{\text{max} }\,\,\,\forall f\in\mathcal{F},\\&\label{Opti_probc}\hspace{.45cm} C_2:
		\hspace{.008cm}\sum_{m\in \mathcal{M}_f}\rho^f_{m,n}\le \ell,\,\,\,\forall n\in\mathcal{N},f\in\mathcal{F},\\&\label{Opti_probe}
		\hspace{.45cm} C_3:
		\hspace{.008cm}\rho^f_{m,n}\in
		\begin{Bmatrix}
			0 ,
			1
		\end{Bmatrix},\,\,\forall m\in \mathcal{M}_f ,n\in\mathcal{N},f\in\mathcal{F},\\&\hspace{.45cm}
		C_4:\hspace{.005cm} p^f_{m,n}\ge0 ,\,\,\, \forall m\in \mathcal{M}_f ,n\in\mathcal{N}, f\in\mathcal{F}, 	
		\\&\label{Opti_probf}
		\hspace{.45cm} C_5:
		\hspace{.005cm} \rho_{m,n}^{f}\rho_{i,n}^{f}Q_{m,i,n}^{f}( \boldsymbol{\rho}, \textbf{p}) \leq 0, \forall f \in \mathcal{F},\nonumber \\
		& \hspace{.45cm}  n \in \mathcal{N}, m,i \in \mathcal{M}_f, \vert h_{i,n}^{f}\vert^2 \leq \vert h _{m,n}^{f}\vert^2, i \neq m,
		\\&\label{Opti_probg}
		\hspace{.45cm} C_6:
		\hspace{.005cm} \rho_{m,n}^{f}\rho_{i,n}^{f}\psi_{m,i,n,e}^{f}( \boldsymbol{\rho}, \textbf{p}) \geq 0, \forall f \in \mathcal{F}, n \in \mathcal{N},\nonumber \\ 
		& \hspace{.45cm}  m,i \in \mathcal{M}_f,    e\in \mathcal{E},  \vert h_{i,n}^{f}\vert^2 \leq \vert h _{e,n}^{f}\vert^2, i \neq m.
	\end{align}
\end{subequations}

The optimization variables  $\mathbf{P}$ and $\boldsymbol{\rho}$ are defined as
$\mathbf{P}
= \left[ {p_{m,n}^f} \right]$ and $\boldsymbol{\rho}=\left[ {\rho_{m,n}^f} \right]
\forall m \in \mathcal{M}_f, n\in \mathcal{N}, f\in \mathcal{F}$, moreover $p^f_{\text{max} }$ is the maximum allowable transmit power at BS $f$.
Constraint $C_1$ demonstrates the maximum allowable transmit power
of BS $f$.  In order to guarantee each subcarrier can be allocated to at most  $\ell$ users, constraint $C_2$ is imposed.  Constraint $C_4$ denotes that the transmit power is non-negative. Constraint $C_5$ guarantees user $m$ can perform SIC successfully on users that $\vert h_{i,n}^{f}\vert^2 \leq \vert h _{m,n}^{f}\vert^2$. Constraint $C_6$ assures  that eavesdropper $e$ is not able to perform SIC, and  other users' signals are treated as interference.

\subsection{Imperfect CSI Scenario}\label{Imperfect CSI}

In this subsection, we assume  imperfect CSI of eavesdroppers is available in the BSs. In particular, the BSs have the knowledge of an
estimated version of channel  i.e., $\hat h_{e,n}^f$ and the channel error is
defined as $e_{h_{e,n}^f} = \tilde h_{e,n}^f -\hat  h_{e,n}^f$. We assume that the channel mismatches lie in the bounded set, i.e., 
$\mathbb{E}_{h_{E}}=\left\{ {{e_{h_{e,n}^f}}: {\left| {{e_{h_{e,n}^f}}} \right|^2} \le  \epsilon} \right\}$ $\forall m\in \mathcal{M}_f ,n\in\mathcal{N},f\in\mathcal{F}$, where ${\epsilon}$ is a known  constant, \cite{Huang}.  Therefore, we model the channel coefficient from BS  $f$ to the $e^{th}$ eavesdropper on subcarrier  $n$ as follows: 
\begin{align}
	{\left|  {\tilde h_{e,n}^f} \right|^2} = & {\left| {\hat h_{e,n}^f + {e_{h_{e,n}^f}}} \right|^2}.
\end{align}
We focus on optimizing the worst-case performance, where
we maximize the worst case sum secrecy rate for the worst channel mismatch $e_{h_{e,n}^f}$ in the bounded set $\mathbb{E}_{h_{E}}$. Hence, the imperfect CSI optimization problem can be formulated as follows: 

\begin{subequations}\label{Opti_prob_imp}
	\begin{align}
	\,\,\,\,\,\,\,\,	&\hspace{-1cm} \max_{\mathbf{P},\boldsymbol{\rho}}\; \min_{\boldsymbol{\varepsilon}}\; \sum_{\forall f\in\mathcal{F}}\sum_{\forall m\in \mathcal{M}_f}\sum_{\forall n\in \mathcal{N}}\rho^f_{m,n} \left\{ {r_{m,n}^f - r_{e_{\max} ,n}^{f,m}} \right\},\\& \label{Opti_probb_imp}
		\hspace{-.5cm}\text{s.t.}:\hspace{.18cm} C_1:\hspace{.002cm}
		\sum_{m\in \mathcal{M}_f}\sum_{n\in \mathcal{N}}\rho^f_{m,n}p^f_{m,n}\le p^f_{\text{max} }\,\,\,\forall f\in\mathcal{F},\\&\label{Opti_probc_imp}\hspace{.45cm} C_2:
		\hspace{.008cm}\sum_{m\in \mathcal{M}_f}\rho^f_{m,n}\le \ell,\,\,\,\forall n\in\mathcal{N},f\in\mathcal{F},\\&\label{Opti_probe_imp}
		\hspace{.45cm} C_3:
		\hspace{.008cm}\rho^f_{m,n}\in
		\begin{Bmatrix}
			0 ,
			1
		\end{Bmatrix},\,\,\forall m\in \mathcal{M}_f ,n\in\mathcal{N},f\in\mathcal{F},\\&\hspace{.45cm}
		C_4:\hspace{.005cm} p^f_{m,n}\ge0 ,\,\,\, \forall m\in \mathcal{M}_f ,n\in\mathcal{N}, f\in\mathcal{F}, 	
		\\&\label{Opti_probf_imp}
		\hspace{.45cm} C_5:
		\hspace{.005cm} \rho_{m,n}^{f}\rho_{i,n}^{f}Q_{m,i,n}^{f}( \boldsymbol{\rho}, \textbf{p}) \leq 0, \forall f \in \mathcal{F},\nonumber \\
		& \hspace{.45cm}  n \in \mathcal{N}, m,i \in \mathcal{M}_f, \vert h_{i,n}^{f}\vert^2 \leq \vert h _{m,n}^{f}\vert^2, i \neq m,
		\\&\label{Opti_probg_imp}
		\hspace{.45cm} C_6:
		\hspace{.005cm} \rho_{m,n}^{f}\rho_{i,n}^{f}\psi_{m,i,n,e}^{f}( \boldsymbol{\rho}, \textbf{p}) \geq 0, \forall f \in \mathcal{F}, n \in \mathcal{N},\nonumber \\ 
		& \hspace{.45cm}  m,i \in \mathcal{M}_f,    e\in \mathcal{E},  \vert h_{i,n}^{f}\vert^2 \leq \vert h _{e,n}^{f}\vert^2, i \neq m,
		\\&\label{Opti_probd_im}\hspace{.45cm} C_7:
		\hspace{.008cm} {\left| {{e_{h_{e,n}^f}}} \right|^2} \le \epsilon, \forall f \in \mathcal{F}, n \in \mathcal{N}, e \in \mathcal{E}
	\end{align}
\end{subequations}
where the optimization variable  $\boldsymbol{\varepsilon}$ is defined as
$\boldsymbol{\varepsilon}=\left[ e_{h_{e,n}^f} \right],
\forall e \in \mathcal{E}, n\in \mathcal{N}, f\in \mathcal{F}$. 
\section{ SOLUTIONS OF THE OPTIMIZATION PROBLEM} \label{SOLUTIONS OF THE OPTIMIZATION PROBLEM}
The optimization problems \eqref{Opti_prob} and \eqref{Opti_prob_imp} are non-convex because they have binary  and continuous variables for subcarrier and power allocation, respectively. Besides, the objective functions are non-convex. Hence, we can not employ existing convex optimization methods straightly. Hence, to tackle this issue, we adopt the well-known  alternative method \cite{T. Wang}, to solve the optimization problems.

\subsection{Solution of the optimization problem in perfect CSI Scenario}\label{solution_perfect}

As there is the $\max$ operator in the objective function, we use a slack variable $\upsilon_{m,n}^{f}$ which is defined as
\begin{align}
	{\max _{e \in \varepsilon }}\left\{ {\log (1 + \gamma _{e,n}^{f,m})} \right\} = {\upsilon_{m,n}^{f}},
\end{align}
by applying  the epigraph method, the optimization problem \eqref{Opti_prob} is rewritten as

\begin{subequations}\label{epi}
	\begin{align}\label{epi_a}
		&\max_{\mathbf{P},\boldsymbol{\rho},\boldsymbol{\upsilon}}\; \sum_{\forall f\in\mathcal{F}}\sum_{\forall m\in \mathcal{M}_f}\sum_{\forall n\in \mathcal{N}} \rho^f_{m,n} \left\{ {r_{m,n}^f -\upsilon_{m,n}^{f} } \right\},\\& 
		\hspace{-.1cm}\text{s.t.}:\hspace{1cm} C_1-C_6, \nonumber
		\\&\label{epi_c}\hspace{.45cm} C^\prime_7:
		\hspace{.008cm} \log \left( {1 + \gamma _{e,n}^{f,m}} \right)\le \upsilon_{m,n}^{f} \,\,\, \forall e\in \mathcal{E}, n\in\mathcal{N},\nonumber\\& \hspace{.45cm} f\in\mathcal{F}, \,\,\, \forall m\in \mathcal{M}_f,
	\end{align}
\end{subequations}
where  $\boldsymbol{\upsilon}$ is defined as
$\boldsymbol{\upsilon}
= \left[ \upsilon_{m,n}^{f} \right],
\forall m \in \mathcal{M}_f, n\in \mathcal{N}, f\in \mathcal{F}$.
For solving \eqref{epi}, we  adopt the well-known  iterative algorithm called ASM. In this method, the optimization problems are converted to two subproblems which one of them has binary  and another has  continuous optimization variables, in  other words, power and subcarrier are allocated
 alternatively \cite{K. Son}. In this method, in each iteration, we allocate transmit power and subcarriers, separately. In other words, in this iterative method,  in each iteration we consider fixed subcarrier  and allocate power, then, for subcarrier allocation we consider fixed power. We summarize the explained algorithm in Algorithm \ref{alg1}.
As seen in this algorithm, it is ended when the stopping condition is satisfied i.e., $\left\| {{\bf{P}}\left( {\mu  + 1} \right) - {\bf{P}}\left( \mu  \right)} \right\| \le \Theta  $, where $ \mu$ and $\Theta$ are the iteration number and stopping threshold, respectively.
\subsubsection{Initialization Method} 
In order to begin the algorithm, we need to initial vectors $\mathbf{P}$  and $\boldsymbol{\rho}$. For initialization, it is supposed that the SBSs do not transmit data, i.e., SBSs at initialization do not serve any users \cite{n.mok, Trong}. In other words, $p_{m,n}^f = 0\,\,\forall f \in {\cal F}/\{ 0\} $. Moreover, 
the subcarriers are allocated to one MBS user that has the highest secrecy rate.
\subsubsection{Subcarrier Allocation}
The subproblem for subcarrier allocation with fixed transmit power (which are computed in the previous iteration) is expressed as:
\begin{align}\label{sub_a}
	&\max_{ \boldsymbol{\rho}}\; \sum_{\forall f\in\mathcal{F}}\sum_{\forall m\in \mathcal{M}_f}\sum_{\forall n\in \mathcal{N}} \rho^f_{m,n} \left\{ {r_{m,n}^f -\upsilon_{m,n}^{f} } \right\},\nonumber\\&  
	\hspace{-.09cm}\text{s.t.}:\hspace{.7cm} C_1-C_3,
	\\&\hspace{1.3cm}C_5-C^\prime_7\nonumber.
\end{align}
Since this optimization problem is Integer Nonlinear Programming (INLP),  we  can solve it by exploiting MADS, to this end, we employ the   NOMAD solver \cite{http}.
\subsubsection{Power Allocation}
\begin{algorithm}[t]
	\caption{ITERATIVE RESOURCE ALLOCATION ALGORITHM For PERFECT CSI} \label{alg1}
	\begin{algorithmic}[1]
		\STATE  \nonumber
		Reformulate the optimization problem via the epigraph method
		\STATE  \nonumber
		Initialization: Set $\mu =0
		\left( {\mu \text{\hspace{.2cm}is the iteration number}} \right)$
		and initialize to $\boldsymbol{\rho}(0)$ and ${\bf{P}}(0)$.
		\STATE \label{set}
		Set $\boldsymbol{\rho}=\boldsymbol{\rho}\left( \mu  \right)$,
		\STATE  		
		Solve \eqref{end} and set  the result  to ${\bf{P}}\left( \mu +1  \right)$,
		\STATE 
		Solve \eqref{sub_a}  and set  the result  to $\boldsymbol{\rho}\left( \mu +1  \right)$,
		\STATE 
		If $\left\| {{\bf{P}}\left( {\mu  + 1} \right) - {\bf{P}}\left( \mu  \right)} \right\| \le \Theta  $\\
		stop,\\
		else\\
		set $\mu = \mu + 1$ and go back to step \ref{set}
	\end{algorithmic}
\end{algorithm}	
The power allocation  subproblem at each iteration when the subcarriers allocation variables are fixed, is expressed as
\begin{align}\label{p_a}
	&\max_{\mathbf{P}, \boldsymbol{\upsilon}}\; \sum_{\forall f\in\mathcal{F}}\sum_{\forall m\in \mathcal{M}_f}\sum_{\forall n\in \mathcal{N}}\rho^f_{m,n} \left\{ {r_{m,n}^f -\upsilon_{m,n}^{f} } \right\},\nonumber\\&  
	\hspace{-.09cm}\text{s.t.}:\hspace{.7cm} C_1, C_4-C^\prime_7.
\end{align}
This optimization subproblem is non-convex because constrains $C^\prime_7$ is non-convex and the objective function is non-concave. To tackle this difficulty, we utilize the SCA approach to approximate constraints $C^\prime_7$ and the objective function. 

First, we investigate $C^\prime_7$:
\begin{align}\label{C_8}
	C^\prime_7: \log \left( {1 + \gamma _{e,n}^{f,m}} \right)\le \upsilon_{m,n}^{f},
\end{align}
by substitution \eqref{SINR_e} into  \eqref{C_8}, we have
\begin{align}\label{c8p}
&	\log \left( {1 + \frac{{p_{m,n}^f{{\left| {h_{e,n}^f} \right|}^2}}}{{{{\left| {h_{e,n}^f} \right|}^2}\sum\limits_{
					\scriptstyle{i\in \mathcal{M}_f/\left\{ m \right\}}\hfill}^{} {p_{i,n}^f\rho _{i,n}^f}  + I_{e,n}^f+{\sigma^2}}}} \right)\\& - \upsilon_{m,n}^{f}\le0, \nonumber
\end{align}
the left hand side of  \eqref{c8p}  is written as follows:
\begin{align}\label{DC81}
	&\log \left( {{{\left| {h_{e,n}^f} \right|}^2}\sum\limits_{i \in {{\cal M}_f}/\left\{ m \right\}}^{} {p_{i,n}^f\rho _{i,n}^f}  + I_{e,n}^f+{\sigma^2} + p_{m,n}^f{{\left| {h_{e,n}^f} \right|}^2}} \right)\nonumber\\& -\log \left( {{{\left| {h_{e,n}^f} \right|}^2}\sum\limits_{i \in {{\cal M}_f}/\left\{ m \right\}}^{} {p_{i,n}^f\rho _{i,n}^f}  + I_{e,n}^f+ {\sigma^2}} \right)-\upsilon_{m,n}^{f}.
\end{align}
As seen, \eqref{DC81} is  the difference between  two concave functions. Hence, we can employ  the  
DC method to approximate \eqref{DC81} to a convex constraint.
To this end, we write \eqref{DC81} as follows:
\begin{align}
	\Xi_{e,n}^{f,m} \left( {\bf{P}} \right) = \Im_{e,n}^{f,m} \left( {\bf{P}} \right) - \Phi_{e,n}^{f,m} ({\bf{P}}),
\end{align}
where
\begin{align}
	&\Im_{e,n}^{f,m}\left( {\bf{P}} \right) \nonumber  =\\&  - \log \left( {{{\left| {h_{e,n}^f} \right|}^2}\sum\limits_{i \in {{\cal M}_f}/\left\{ m \right\}}^{} {p_{i,n}^f\rho _{i,n}^f}  +I_{e,n}^f+ {\sigma^2}} \right) -\upsilon_{m,n}^{f},
\end{align}
and
\begin{align}
	&	\Phi_{e,n}^{f,m}\left( {\bf{P}} \right)  \nonumber =- \\& \log \left( {{{\left| {h_{e,n}^f} \right|}^2}\sum\limits_{i \in {{\cal M}_f}/\left\{ m \right\}}^{} {p_{i,n}^f\rho _{i,n}^f}  + p_{m,n}^f{{\left| {h_{e,n}^f} \right|}^2} + I_{e,n}^f+{\sigma^2}} \right)
\end{align}
$\Im_{e,n}^{f,m}\left( {\bf{P}} \right)$ and $\Phi_{e,n}^{f,m}\left( {\bf{P}} \right)$ are convex,  by utilizing a linear approximation,   $\Phi_{e,n}^{f,m}\left( {\bf{P}} \right)$ can be written as follows:
\begin{align}
	& \Phi_{e,n}^{f,m}\left( {{\bf{P}}} \right)\simeq \tilde \Phi_{e,n}^{f,m}\left( {{\bf{P}}} \right) =	\Phi_{e,n}^{f,m}\left( {{\bf{P}}\left( {\mu  - 1} \right)} \right) + \\&{\nabla ^T}\Phi_{e,n}^{f,m}\left( {{\bf{P}}\left( {\mu  - 1} \right)} \right)\nonumber\left( {{\bf{P}} - {\bf{P}}(\mu  - 1)} \right),
\end{align}
where ${\nabla}\Phi_{e,n}^{f,m}\left( {{\bf{P}}} \right)$, is the gradient of $\Phi_{e,n}^{f,m}\left( {{\bf{P}}} \right)$ which is defined as:
\begin{align}\label{grad0}
	&{\nabla}\Phi_{e,n}^{f,m} \left( {\bf{P}} \right) = \frac{\partial}{{\partial{\bf{P}}}}\Phi_{e,n}^{f,m} \left( {\bf{P}} \right) =\\& \left[ {\frac{{\partial \Phi_{e,n}^{f,m}\left( {\bf{P}} \right) }}{{\partial p_{m,n}^f}}} \right],\forall m \in \mathcal{{M}}_f,\forall n \in \mathcal{N}, \forall f \in \mathcal{F}, \forall e\in \mathcal{E}, \nonumber
\end{align}
and
\begin{align}
	&{\frac{{\partial \Phi_{e,n}^{f,m}\left( {\bf{P}} \right) }}{{\partial p_{a,b}^c}}}=\begin{array}{l}
		\left\{ {\begin{array}{*{20}{l}}
				{X,}& a=m, b=n, c=f,\\
				{ Y,}&{\forall a \in {\mathcal{M}_f}/\left\{ m \right\}, b=n, c=f,}
			\\
			{ B,}&{\forall a \in {\mathcal{M}_{f^\prime}}, b=n, c=f^\prime\in \mathcal{F}/f,}
				\\
				{ 0,}&{O.W,}
		\end{array}} \right.
	\end{array}
\end{align}
%############################3
moreover,  $X$, $Y$,  and $B$ are calculated as follows:
\begin{align}
X = - \frac{{{{\left| {h_{e,n}^f} \right|}^2}}}{{{{\left| {h_{e,n}^f} \right|}^2}\sum\limits_{i \in {{\cal M}_f}/\left\{ m \right\}}^{} {p_{i,n}^f\rho _{i,n}^f}  + p_{m,n}^f{{\left| {h_{e,n}^f} \right|}^2} +I_{e,n}^f+ {\sigma^2}}}, 
\end{align}	
\begin{align}\label{grad01}
	&Y=  - \frac{{{{\left| {h_{e,n}^f} \right|}^2}\rho _{a,n}^f }}{{{{\left| {h_{e,n}^f} \right|}^2}\sum\limits_{i \in {{\cal M}_f}/\left\{ m \right\}}^{} {p_{i,n}^f\rho _{i,n}^f}  + p_{m,n}^f{{\left| {h_{e,n}^f} \right|}^2} + I_{e,n}^f+{\sigma^2}}},
	 \end{align} 
	 \begin{align}\label{grad02}
	 &B=  - \frac{{{{\left| {h_{e,n}^c} \right|}^2}\rho _{a,n}^c }}{{{{\left| {h_{e,n}^f} \right|}^2}\sum\limits_{i \in {{\cal M}_f}/\left\{ m \right\}}^{} {p_{i,n}^f\rho _{i,n}^f}  + p_{m,n}^f{{\left| {h_{e,n}^f} \right|}^2} +I_{e,n}^f+ {\sigma^2}}},
	 \end{align}
therefore, $\nabla ^T\Phi_{e,n}^{f,m} \left( {\bf{P}} \right)$ is a vector that its length is $N\times M\times F$. 
After  approximation $C^\prime_7$ to the convex constraint, we convert the  objective function to a concave function by exploiting the DC method, hence we have: 
\begin{align}\label{obj}
	\log (1 + \gamma _{m,n}^f)- \upsilon_{m,n}^{f},
\end{align}
by substitution \eqref{SINR_m} into \eqref{obj}, we have
\begin{align}
	\log (1 +  \frac{{p_{m,n}^f{{\left| {h_{m,n}^f} \right|}^2}}}{{{{\left| {h_{m,n}^f} \right|}^2}\sum\limits_{{{\left| {h_{m,n}^f} \right|}^2} \le {{\left| {h_{i,n}^f} \right|}^2}\hfill\atop
				\scriptstyle i \in {\mathcal{M}_f}/\left\{ m \right\}\hfill}^{} {p_{i,n}^f\rho _{i,n}^f}  +I_{m,n}^f+ {\sigma^2}}})- \upsilon_{m,n}^{f},
\end{align}
where can be written as:
\begin{align}\label{DC_C9}
	&	\log ( {{\left| {h_{m,n}^f} \right|}^2}\sum\limits_{{{\left| {h_{m,n}^f} \right|}^2} \le {{\left| {h_{i,n}^f} \right|}^2}\hfill\atop
			\scriptstyle i \in {\mathcal{M}_f}/\left\{ m \right\}\hfill}^{} {p_{i,n}^f\rho _{i,n}^f}  + I_{m,n}^f+{\sigma^2} + \nonumber\\& p_{m,n}^f{{\left| {h_{m,n}^f} \right|}^2})  - \log( {{\left| {h_{m,n}^f} \right|}^2}\sum\limits_{{{\left| {h_{m,n}^f} \right|}^2} \le {{\left| {h_{i,n}^f} \right|}^2}\hfill\atop
			\scriptstyle i\in {\mathcal{M}_f}/\left\{ m \right\}\hfill}^{} {p_{i,n}^f\rho _{i,n}^f}  +I_{m,n}^f+\nonumber\\ &  {\sigma^2} ) - \upsilon_{m,n}^{f},\end{align}
we can writhe \eqref{DC_C9} as follows:
\begin{align}
	{{ U_{m,n}^f}}\left( {\bf{P}} \right){{  =  G_{m,n}^f}}\left( {\bf{P}} \right){{  -  H_{m,n}^f}}\left( {\bf{P}} \right),
\end{align}
where ${{H_{m,n}^f}}\left( {\bf{P}} \right)$ and ${{G_{m,n}^f}}\left( {\bf{P}} \right)$  are defined as
%\begin{align}
%	&{{G_{m,n}^f}}\left( {\bf{P}} \right)= -\upsilon_{m,n}^{f}+  \log \left( {{{\left| {h_{m,n}^f} \right|}^2}} \right.\\&\left. {\sum\limits_{\begin{array}{*{20}{c}}
%			{{{\left| {h_{m,n}^f} \right|}^2} \le {{\left| {h_{i,n}^f} \right|}^2}}\\
%			{i \in {{\cal M}_f}/\left\{ m \right\}}
%			\end{array}}^{} {p_{i,n}^f\rho _{i,n}^f}  + p_{m,n}^f{{\left| {h_{m,n}^f} \right|}^2} + I_{m,n}^f + {\sigma ^2}} \right),\nonumber
%\end{align} 
\begin{align}
&{{G_{m,n}^f}}\left( {\bf{P}} \right)= -\upsilon_{m,n}^{f}+ \log	\\&  ( {{{{\left| {h_{m,n}^f} \right|}^2}}\sum\limits_{{{\left| {h_{m,n}^f} \right|}^2} \le {{\left| {h_{i,n}^f} \right|}^2}\hfill\atop
		\scriptstyle i\in {\mathcal{M}_f}/\left\{ m \right\}\hfill}^{} {p_{i,n}^f\rho _{i,n}^f}  + p_{m,n}^f{{\left| {h_{m,n}^f} \right|}^2} + I_{m,n}^f + {\sigma ^2}} ),\nonumber
\end{align}
and
\begin{align}\label{H}
	&{{H_{m,n}^f}}\left( {\bf{P}} \right) =	\\&  \log \left( {{{\left| {h_{m,n}^f} \right|}^2}\sum\limits_{{{\left| {h_{m,n}^f} \right|}^2} \le {{\left| {h_{i,n}^f} \right|}^2}\hfill\atop
			\scriptstyle i\in {\mathcal{M}_f}/\left\{ m \right\}\hfill}^{} {p_{i,n}^f\rho _{i,n}^f}  + I_{m,n}^f+{\sigma^2}} \right),\nonumber
\end{align}
respectively. $G_{m,n}^f\left( {\bf{P}} \right)$ and $H_{m,n}^f\left( {\bf{P}} \right)$ are concave,  by utilizing a linear approximation we can write $H_{m,n}^f\left( {\bf{P}} \right)$ as follows:
\begin{align}\label{aprox}
	&	H_{m,n}^f\left( {{\bf{P}}} \right)\simeq \tilde H_{m,n}^f\left( {{\bf{P}}} \right) = H_{m,n}^f\left( {{\bf{P}}\left( {\mu  - 1} \right)} \right) + \\&{\nabla ^T}H_{m,n}^f\left( {{\bf{P}}\left( {\mu - 1} \right)} \right)\nonumber\left( {{\bf{P}} - {\bf{P}}(\mu  - 1)} \right),
\end{align}
%$
where ${\nabla ^T}H_{m,n}^f\left( {{\bf{P}}} \right)$ is calculated as follows:
\begin{align}\label{grad}
	&{\nabla ^T}H_{m,n}^f \left( {\bf{P}} \right) = \frac{\partial}{{\partial{\bf{P}}}}H_{m,n}^f \left( {\bf{P}} \right) \\& =\left[ {\frac{{\partial H_{m,n}^f\left( {\bf{P}} \right) }}{{\partial p_{m,n}^f}}} \right],\forall m \in \mathcal{{M}}_f,\forall n \in \mathcal{N},\forall f \in \mathcal{F}, \nonumber
\end{align}
%#############################
%		\begin{align}
%	\begin{array}{l}
%	{\nabla ^T}H_{m,n}^f\left( {{\bf{P}}} \right) = 
%	\left\{ {\begin{array}{*{20}{l}}
%		{X}&{{{\left| {h_{i,n}^f} \right|}^2} \le {{\left| {h_{m,n}^f} \right|}^2}},\\ {} & a=m, b=n, c=f\\
%		{ Y}&{O.W}
%		\end{array}} \right.
%	\end{array}
%	\end{align}
We take derivative of $H_{m,n}^f \left( {\bf{P}} \right) $ with respect to $p_{a,b}^c$ as follows:
%###############################
%###############################
\begin{align}
	&{\frac{{\partial H_{m,n}^f\left( {\bf{P}} \right) }}{{\partial p_{a,b}^c}}}=\\&\begin{array}{l}
		\left\{ {\begin{array}{*{20}{l}}
				{ Z} &{\forall a \in {\mathcal{M}_f}/\left\{ m \right\}, b=n, c=f, {{\left| {h_{m,n}^f} \right|}^2} \le {{\left| {h_{a,n}^f} \right|}^2},}
				\\
				{ T}&{\forall a \in {\mathcal{M}_{f^\prime}}, b=n, c=f^\prime\in \mathcal{F}/f}
				\\
				{ 0}&{O.W}
		\end{array}} \right.
	\end{array}\nonumber
\end{align}

where  $Z$ and $T$  are calculated as follows:
\begin{align}\label{grad1}
	&Z =\\&   \frac{{{{\left| {h_{m,n}^f} \right|}^2}\rho _{a,n}^f }}{{{\left| {h_{m,n}^f} \right|}^2}\sum\limits_{{{\left| {h_{m,n}^f} \right|}^2} \le {{\left| {h_{i,n}^f} \right|}^2}\hfill\atop
			\scriptstyle i\in {\mathcal{M}_f}/\left\{ m \right\}\hfill}^{} {p_{i,n}^f\rho _{i,n}^f}  + I_{m,n}^f + {\sigma^2}},\nonumber
\end{align} 
\begin{align}\label{grad11}
&T =\\&   \frac{{{{\left| {h_{m,n}^c} \right|}^2}\rho _{a,n}^c }}{{{\left| {h_{m,n}^f} \right|}^2}\sum\limits_{{{\left| {h_{m,n}^f} \right|}^2} \le {{\left| {h_{i,n}^f} \right|}^2}\hfill\atop
		\scriptstyle i\in {\mathcal{M}_f}/\left\{ m \right\}\hfill}^{} {p_{i,n}^f\rho _{i,n}^f}  + I_{m,n}^f+ {\sigma^2}}.\nonumber
\end{align} 
Consequently, we have a 
convex optimization problem in the canonical form, by exploiting the DC approximation, which
is formulated as:

\begin{subequations}\label{end}
	\begin{align}\label{end_a}
		&\max_{\mathbf{P},\boldsymbol{\upsilon}}\; \sum_{\forall f\in\mathcal{F}}\sum_{\forall m\in \mathcal{M}_f}\sum_{\forall n\in \mathcal{N}} \rho^f_{m,n} \left\{ {G_{m,n}^f\left( {\bf{P}} \right) - \tilde H_{m,n}^f\left( {\bf{P}} \right)} \right\},\\& \label{end_b}
		\hspace{-.1cm}\text{s.t.}:\hspace{1cm} C_1-C_7,
		\\&\label{end_c}\hspace{.45cm}C^{\prime}_7:
		\hspace{.008cm}\Im_{e,n}^{f,m} \left( {\bf{P}} \right) - \tilde{\Phi}_{e,n}^{f,m} ({\bf{P}})\le0 \,\,\, \forall e\in \mathcal{E},\\& \hspace{.45cm} n\in\mathcal{N}, f\in\mathcal{F}, m\in\mathcal{M}_f, \nonumber
	\end{align}
\end{subequations}
For solving the  convex optimization problem \eqref{end}, we can use available softwares, such  as CVX solver \cite{CVX}.
%	\vspace{1cm}
\subsubsection{Convergence of the algorithm}
In this subsection,  we prove the convergence of the algorithm and illustrate that after each iteration the value of objective function $f{\rm{ }}\left( {{\boldsymbol{\rho }},{\rm{ }}{\bf{p}}} \right) = \rho _{m,n}^f\left\{ {r_{m,n}^f - \upsilon _{e,n}^{f,m}} \right\}$, is improved  and converged.

\textit{Proof}: In this algorithm, after applying the third step, with a given ${\boldsymbol{\rho }} = {\boldsymbol{\rho }}\left( \mu  \right)$, the power allocation of iteration $\mu  + 1$  is obtained. According to Appendix I, we will have $f \left( {{\boldsymbol{\rho }} \left( \mu  \right), {\bf{p}}\left( \mu  \right)} \right){\rm{ }} \le {\rm{ }}f\left( {{\rm{ }}{\boldsymbol{\rho }}\left( \mu  \right),{\rm{ }}{\bf{p}}\left( {\mu  + 1} \right){\rm{ }}} \right)$. Moreover, in the fourth step,  with a given ${\bf{p}} = {\bf{p}}\left( {\mu  + 1} \right)$,  the subcarrier allocation of  this iteration   is obtained. According to this fact that, after each iteration, subcarrier allocation with  feasible power solution  improves the objective function, hence we have:
\begin{align}
	... \le f\left( {{\rm{ }}{\boldsymbol{\rho }}\left( \mu  \right),{\rm{ }}{\bf{p}}\left( \mu  \right){\rm{ }}} \right){\rm{ }} \le {\rm{ }}f\left( {{\rm{ }}{\boldsymbol{\rho }}\left( \mu  \right),{\rm{ }}{\bf{p}}\left( {\mu  + 1} \right){\rm{ }}} \right) \le\\ f\left( {{\rm{ }}{\boldsymbol{\rho }}\left( {\mu  + 1} \right),{\rm{ }}{\bf{p}}\left( {\mu  + 1} \right){\rm{ }}} \right) \le ... \le f\left( {{\rm{ }}{{\boldsymbol{\rho }}^*},{\rm{ }}{{\bf{p}}^*}{\rm{ }}} \right)\nonumber
\end{align}
where ${{\boldsymbol{\rho }}^*}$ and ${{\bf{p}}^*}$ are obtained at the last iteration, \cite{R8}. Convergence behavior of the proposed algorithm is shown in Fig. \ref{itr}.   It should be noted, globally optimal solution is not guaranteed by this solution even after convergence. Hence, for finding the globally optimal solution, we utilize the monotonic
optimization method which is explained,   in Section \ref{Optimal solution}.
\subsubsection{Computational complexity}
Solution of the optimization problem \eqref{Opti_prob} consists of two stages 1) Calculation of  power allocation from problem \eqref{end}, 2) calculation of subcarrier allocation from problem \eqref{sub_a}. As we know, CVX software employs geometric programing with the Interior Point Method (IPM) \cite{CVX}, hence, the order of computational complexity can be obtained as:
\begin{align}
	O\left( {\frac{{\log \left( {\frac{{NOC}}{{t\partial }}} \right)}}{{\log \left( \xi  \right)}}} \right),
\end{align}
where NOC is the total number of constraints. $\partial $ , $\xi $ and $t$ are parameters of IPM.  $ 0 \le \partial  <  < 1$ is the stopping criterion of IPM,  $\xi$ is used for the accuracy IPM and 
$t$  is initial point for approximated the accuracy of IPM, \cite{n.mok}, \cite{R5}.   Hence, the complexity order is given by:

\begin{align}
O\left( {\frac{{\log \left( {\frac{{F\left( {1 + N\left( {1 + M + M\left( {M - 1} \right) + ME\left( {M - 1} \right) + ME} \right)} \right)}}{{t\partial }}} \right)}}{{\log \left( \xi  \right)}}} \right)
\end{align}
\subsection{Solution of the optimization problem in Imperfect CSI Scenario}\label{solution_imperfect}

For solving \eqref{Opti_prob_imp}, first we solve the inner minimization and obtain $ \boldsymbol{\varepsilon}$, then solve the maximization problem according to Section \ref{solution_perfect}. The inner minimization can be written as follows:
\begin{subequations}\label{dtt}
	\begin{align} 
		&\min_{\boldsymbol{\varepsilon}}\; \sum_{\forall f\in\mathcal{F}}\sum_{\forall m\in \mathcal{M}_f}\sum_{\forall n\in \mathcal{N}}\rho^f_{m,n} \left\{ {r_{m,n}^f -\upsilon_{m,n}^{f} } \right\}, \\& \text{s.t.}:\hspace{.18cm} C_6, C_7, C^\prime_7.
	\end{align}
\end{subequations}
Our aim is to minimize the objective function, to this end, we should maximize $\upsilon_{m,n}^{f}$. Hence, according to $C_7^{\prime}$, we maximize lower bound of  $\upsilon_{m,n}^{f}$, i.e., $\log \left( {1 + \gamma _{e,n}^{f,m}} \right)$. Since the logarithmic function is increasing,  we can maximize  $\gamma _{e,n}^{f,m}$ instead of $\log \left( {1 + \gamma _{e,n}^{f,m}} \right)$. As $\gamma _{e,n}^{f,m}$ is fractional, we should maximize numerator and minimize denominator. To this end, we use the triangle inequality which is defined as follows:
\begin{align}\label{Tr_enq}
&{\left| {\tilde h_{e,n}^f} \right|^2} -\epsilon \le {\left| {\tilde h_{e,n}^f} \right|^2} - {\left| {{e_{h_{e,n}^f}}} \right|^2} \le {\left| {\hat h_{e,n}^f + {e_{h_{e,n}^f}}} \right|^2} \le \nonumber\\&
 {\left| {\tilde h_{e,n}^f} \right|^2} + {\left| {{e_{h_{e,n}^f}}} \right|^2} \le {\left| {\tilde h_{e,n}^f} \right|^2}+\epsilon.
\end{align}
By using \eqref{Tr_enq}, we can write the upper bound of  $\gamma _{e,n}^{f,m}$ as follows:
\begin{align}\label{UP_bound}
& \frac{{p_{m,n}^f{{\left| {h_{e,n}^f} \right|}^2}}}{{{{\left| {h_{e,n}^f} \right|}^2}\sum\limits_{
			\scriptstyle{i\in \mathcal{M}_f/\left\{ m \right\}}\hfill}^{} {p_{i,n}^f\rho _{i,n}^f}  + I_{e,n}^f+{\sigma^2}}} \le \nonumber \\&
 \frac{{p_{m,n}^f} \left({\left| {\tilde h_{e,n}^f} \right|^2} + \epsilon\right)} {{{ \left({\left| {\tilde h_{e,n}^f} \right|^2} - \epsilon\right) }\sum\limits_{
			\scriptstyle{i\in \mathcal{M}_f/\left\{ m \right\}}\hfill}^{} {p_{i,n}^f\rho _{i,n}^f}  + \tilde I_{e,n}^f+{\sigma^2}}}.
\end{align}
where $\tilde I_{e,n}^f = \sum\limits_{{f^\prime } \in {\cal F}/f}^{} \left({{\left| {{\tilde h_{e,n}^{f'}}} \right|}^2} - \epsilon\right) \sum\limits_{i \in {{\cal M}_{{f^\prime }}}}^{} {\rho _{i,n}^{f'}p_{i,n}^{f'}} $. 
Also, we consider the worst case for constraint $C_6$. According to \eqref{Tr_enq}, we can rewrite the worst case $C_6$ as follows: 
\begin{align}
&\Psi _{m,i,n,e}^f({\boldsymbol{\rho }},{\bf{p}}) \ge \tilde \Psi _{m,i,n,e}^f({\boldsymbol{\rho }},{\bf{p}})= |h_{i,n}^f{|^2}{\sigma^2}+ |h_{i,n}^f{|^2}{\tilde I_{e,n}^f} -\nonumber \\&
\left({\left| {\tilde h_{e,n}^f} \right|^2} + \epsilon\right) \left({{\left| {h_{i,n}^f} \right|^2}\sum\limits_{{{\left| {h_{i,n}^f} \right|}^2} \le {{\left| {h_{l,n}^f} \right|}^2}\hfill\atop
		\scriptstyle l \in {\mathcal{M}_f}/\left\{ i \right\}\hfill}^{} {p_{l,n}^f\rho _{l,n}^f}+I_{i,n}^f+\sigma^2}\right)                
  +\nonumber \\&\left({\left| {\tilde h_{e,n}^f} \right|^2} - \epsilon\right)
{\left| {h_{i,n}^f} \right|^2}\sum\limits_{l \in {\mathcal{M}_f}/\left\{ i \right\}}^{} {p_{l,n}^f\rho _{l,n}^f}  \ge 0.
\end{align}

In the following, we should solve the outer maximization, which is written as follows:
\begin{subequations}\label{p_rho}
\begin{align}
&\max_{\mathbf{P},\boldsymbol{\rho},\boldsymbol{\upsilon}}\; \sum_{\forall f\in\mathcal{F}}\sum_{\forall m\in \mathcal{M}_f}\sum_{\forall n\in \mathcal{N}}\rho^f_{m,n} \left\{ {r_{m,n}^f -\upsilon_{m,n}^{f} } \right\},\\&  
\hspace{-.09cm}\text{s.t.}:\hspace{.7cm} C_1-C_5,
	\\&\label{O}
	\hspace{.45cm} C_6^\prime:
	\hspace{.005cm} \rho_{m,n}^{f}\rho_{i,n}^{f} \tilde \psi_{m,i,n,e}^{f}( \boldsymbol{\rho}, \textbf{p}) \geq 0, \forall f \in \mathcal{F}, n \in \mathcal{N},\nonumber \\ 
	& \hspace{.45cm}  m,i \in \mathcal{M}_f,    e\in \mathcal{E},  \vert h_{i,n}^{f}\vert^2 \leq \vert h _{e,n}^{f}\vert^2, i \neq m,
	\\&\label{0O}
	\hspace{.45cm} C_7^{''}:\\&
		\log \left( {1 + \frac{{p_{m,n}^f} \left({\left| {\tilde h_{e,n}^f} \right|^2} + \epsilon\right)} {{{ \left({\left| {\tilde h_{e,n}^f} \right|^2} - \epsilon\right) }\sum\limits_{
						\scriptstyle{i\in \mathcal{M}_f/\left\{ m \right\}}\hfill}^{} {p_{i,n}^f\rho _{i,n}^f}  +\tilde I_{e,n}^f+ {\sigma^2}}}} \right) \nonumber \\& \hspace{.45cm} -\upsilon_{m,n}^{f}\le 0, \,\,\, \forall e\in \mathcal{E}, n\in\mathcal{N}, f\in\mathcal{F}, \,\,\, \forall m\in \mathcal{M}_f.
\end{align}
\end{subequations}
This optimization problem \eqref{p_rho} can be solved similar to the proposed approach in Section  \ref{solution_perfect}.

\section{Massive Connectivity Scenario}\label{MCS}
In this section, we aim to evaluate  the PD-NOMA technique in ultra dense network for secure massive connectivity in 5G networks. Without loss of generality, for changing our scenario from HetNet to HUDN, we need to extend the dimension of system model. According to \cite{Z.Qin}, \cite{P.Soldati}, and \cite{J.Zhao}, in order to tackle high dimension complexity of resource allocation in HUDNs and overcome hardware computation limitations, it is assumed that the transmit power is uniformly allocated to devices/users and subcarriers are dynamically allocated.  

To know the performance degradation due to the uniform power allocation, we compare the uniform power allocation method for a small network dimension with our proposed solution  i.e., joint power and subcarrier allocation in section of simulation result. Based on simulation results, we show the performance of uniform power allocation is close to the performance of our proposed solution  i.e., joint power and subcarrier allocation in the small network dimension.

\section{Optimal Solution}\label{Optimal solution}
In this section, our aim is to solve problem \eqref{epi} by utilizing  the monotonic optimization method. Hence, in this section, three main steps are  performed  as follows:
\begin{enumerate}
	\item
	Problem \eqref{epi} is transformed into an optimization problem at which its optimization variables are only transmit power.
	\item The new optimization problem is converted to a canonical form of monotonic optimization problem.
	\item
	Finally, by exploiting the polyblock algorithm, we solve the monotonic optimization problem globally.
\end{enumerate}

\subsection{Problem Transformation}
For the  first step, we assume  each subcarrier can be allocated to  at most  two users, i.e., $\ell=2$. Hence, based on constraints \eqref{Opti_probc} and \eqref{Opti_probe},  if $p_{m,n}^f \ne 0$ and $p_{i,n}^f \ne 0$, we have $p_{w,n}^f=0 \,\, \forall m ,i ,w\in \mathcal{M}_f$, $m \ne i \ne w$. Therefore, constraint \eqref{Opti_probc}  is equivalent to 
\begin{align}\label{T1}
&p_{m,n}^fp_{i,n}^fp_{w,n}^f \le 0, \,\,\, \forall  n\in\mathcal{N}, \,\,\, f\in\mathcal{F},  \,\,\,  m ,i, w\in \mathcal{M}_f,\\& \,\,\, m \ne i \ne w \nonumber
\end{align} 
Therefore, the optimization problem \eqref{epi} can be transformed into  a new optimization problem with only transmit power variables as follows:

 \begin{subequations}\label{Opt_prob}
 	\begin{align}
 	&\max_{\mathbf{P},\boldsymbol{\upsilon}}\; \sum_{\forall f\in\mathcal{F}}\sum_{\forall m\in \mathcal{M}_f}\sum_{\forall n\in \mathcal{N}} \left\{ {\tilde{r}_{m,n}^f - \upsilon_{m,n}^{f}} \right\},\\& \label{Opt_probb}
 	\text{s.t.}:\hspace{.18cm} \hspace{.008cm}
 	\sum_{m\in \mathcal{M}_f}\sum_{n\in \mathcal{N}} p^f_{m,n}\le p^f_{\text{max} }\,\,\,\forall f\in\mathcal{F},
 	\\&\label{Opt_probc} \hspace{.45cm}
 	\hspace{.005cm} p^f_{m,n}\ge0 ,\,\,\, \forall m\in \mathcal{M}_f ,n\in\mathcal{N}, m\in\mathcal{M}_f, 	
 	\\&\label{Opt_probd}
 	\hspace{.45cm} 
 	\hspace{.005cm} p_{m,n}^{f} p_{i,n}^{f}\hat{Q}_{m,i,n}^{f} \leq 0, \forall f \in \mathcal{F}, \\
 	& \hspace{.45cm}  n \in \mathcal{N}, m,i \in \mathcal{M}_f, \vert h_{i,n}^{f}\vert^2 \leq \vert h _{m,n}^{f}\vert^2, i \neq m, \nonumber
 	\\&\label{Opt_probe}
 	\hspace{.45cm}
 	\hspace{.005cm} -p_{m,n}^{f}p_{i,n}^{f}\hat{\psi}_{m,i,n,e}^{f} \leq 0, \forall f \in \mathcal{F}, n \in \mathcal{N}, \\ 
 	& \hspace{.45cm}  m,i \in \mathcal{M}_f,    e\in \mathcal{E},  \vert h_{i,n}^{f}\vert^2 \leq \vert h _{e,n}^{f}\vert^2, i \neq m, \nonumber
 	\\&\hspace{.45cm}
 	\hspace{.008cm}\log \left( {{{\left| {h_{e,n}^f} \right|}^2}\sum\limits_{i \in {{\cal M}_f}/\left\{ m \right\}}^{} {p_{i,n}^f}  +\hat I_{m,n}^f+ {\sigma^2} + p_{m,n}^f{{\left| {h_{e,n}^f} \right|}^2}} \right)- \nonumber\\& \hspace{.45cm}\log \left( {{{\left| {h_{e,n}^f} \right|}^2}\sum\limits_{i \in {{\cal M}_f}/\left\{ m \right\}}^{} {p_{i,n}^f}  +\hat I_{m,n}^f+  {\sigma^2}} \right)-\upsilon_{m,n}^{f} \le 0,\nonumber \\& 
 	 \label{Opt_probf} \hspace{.45cm}  \forall e\in \mathcal{E}, n\in\mathcal{N}, f\in\mathcal{F}, \,\,\, \forall m\in \mathcal{M}_f,   \\&\hspace{.45cm} \label{Opt_probg}
 	\hspace{.005cm} p_{m,n}^fp_{i,n}^fp_{w,n}^f \le 0, \,\,\, \forall  n\in\mathcal{N}, \,\,\, f\in\mathcal{F},  \,\,\,  m ,i, w\in \mathcal{M}_f,\nonumber\\& \hspace{.45cm} m \ne i \ne w, 
 	\end{align}
 \end{subequations}
where

\begin{align}
&\hat{Q}_{m,i,n}^{f}= - |h_{m,n}^f{|^2}{\sigma^2} + |h_{i,n}^f{|^2}{\sigma^2} +|h_{i,n}^f{|^2}\hat I_{m,n}^f - |h_{m,n}^f{|^2}\hat I_{i,n}^f\nonumber \\&- {\left| {h_{m,n}^f} \right|^2}{\left| {h_{i,n}^f} \right|^2}\sum\limits_{{{\left| {h_{i,n}^f} \right|}^2} \le {{\left| {h_{l,n}^f} \right|}^2}\hfill\atop
	\scriptstyle l \in {\mathcal{M}_f}/\left\{ i \right\}\hfill}^{} {p_{l,n}^f}  + \nonumber\\& {\left| {h_{i,n}^f} \right|^2}{\left| {h_{m,n}^f} \right|^2}\sum\limits_{{{\left| {h_{m,n}^f} \right|}^2} \le {{\left| {h_{l,n}^f} \right|}^2}\hfill\atop
	\scriptstyle l \in {\mathcal{M}_f}/\left\{i  \right\}\hfill}^{} {p_{l,n}^f}, 
\end{align}
and
\begin{align}
&\hat{\psi}_{m,i,n,e}^{f}=	
- |h_{e,n}^f{|^2}{\sigma^2} + |h_{i,n}^f{|^2}{\sigma^2}- |h_{e,n}^f{|^2}\hat I_{i,n}^f +|h_{i,n}^f{|^2}\hat I_{e,n}^f-\nonumber \\&
{\left| {h_{e,n}^f} \right|^2}{\left| {h_{i,n}^f} \right|^2}\sum\limits_{{{\left| {h_{i,n}^f} \right|}^2} \le {{\left| {h_{l,n}^f} \right|}^2}\hfill\atop
	\scriptstyle l \in {\mathcal{M}_f}/\left\{ i \right\}\hfill}^{} {p_{l,n}^f}  +  \\&
{\left| {h_{i,n}^f} \right|^2}{\left| {h_{e,n}^f} \right|^2}\sum\limits_{l \in {\mathcal{M}_f}/\left\{ i \right\}}^{} {p_{l,n}^f}, \nonumber
\end{align}
where $\hat I_{m,n}^f=\sum\limits_{{f^\prime } \in {\cal F}/f}^{} {{\left| {{h_{m,n}^{f'}}} \right|}^2} \sum\limits_{i \in {{\cal M}_{{f^\prime }}}}^{} {p_{i,n}^{f'}}$ and $\hat I_{e,n}^f = \sum\limits_{{f^\prime } \in {\cal F}/f}^{} {{\left| {{h_{e,n}^{f'}}} \right|}^2} \sum\limits_{i \in {{\cal M}_{{f^\prime }}}}^{} {p_{i,n}^{f'}} $.
Moreover, $\tilde{r}_{m,n}^f$ is defined as: 
%	\begin{table}[b] 
%	\centering
%	\caption{Complexity in the ASM and Monotonic}
%	\begin{tabular}{|m{12 em} |c |c |} 
%		\hline
%		Scenarios & Subcarrier allocation & Power allocation   \\ [0.5ex] 
%		\hline
%		\vspace{.6cm}
%		ASM  & $ O\left( {\frac{{\log \left( {\frac{{F\left( {1 + N\left( {1 + M + M\left( {M - 1} \right) + ME\left( {M - 1} \right) + ME} \right)} \right)}}{{t\partial }}} \right)}}{{\log \left( \xi  \right)}}} \right)$& $O\left( {\frac{{\log \left( {\frac{{F\left( {1 + N\left( {M + M\left( {M - 1} \right) + ME\left( {M - 1} \right) + ME} \right)} \right)}}{{t\partial }}} \right)}}{{\log \left( \xi  \right)}}} \right)$\\ 
%		\hline 
%		\vspace{.6cm}
%		Monotonic  & $O\left( {F + NF\left( {3M + 1} \right) + \left( {M - 1} \right)\left( {M\left( {M - 2} \right) + 3NFM + 3NFME} \right) + 3FNME} \right)$ & $-$\\ 
%		\hline 
%	\end{tabular}
%	\label{table}
%\end{table}

\begin{align}
\tilde{r}_{m,n}^f=\log \left( {1 +\frac{{p_{m,n}^f{{\left| {h_{m,n}^f} \right|}^2}}}{{{{\left| {h_{m,n}^f} \right|}^2}\sum\limits_{\scriptstyle{\left| {h_{m,n}^f} \right|^2} \le {\left| {h_{i,n}^f} \right|^2}\hfill\atop
				\scriptstyle i \in {\mathcal{M}_f}/\left\{ m \right\}\hfill}^{} {p_{i,n}^f}  +\hat I_{m,n}^f+ {\sigma^2}}} } \right),
\end{align}
\subsection{Monotonic Optimization}
In the second step, our aim is to formulate the optimization problem \eqref{Opt_prob} as a monotonic optimization problem in the canonical form. As we know,  the optimization problem \eqref{Opt_prob} is a non-monotonic problem because the objective function is not increasing function and constraints \eqref{Opt_probd},   \eqref{Opt_probe}, and \eqref{Opt_probf} are not inside normal or conormal sets.
Since the  optimization problem is a problem with hidden monotonicity \cite{48}, we can  rewrite the objective function and non-monotonic constraints to a differences of increasing function form. Hence, let us  reformulate the objective function as ${\tilde{r}_{m,n}^f - \upsilon_{m,n}^{f}}=g_{m,n}^{f+} -g_{m,n}^{f-} $ where $g_{m,n}^{f+}=\log \left( {{{\left| {h_{m,n}^f} \right|}^2}\sum\limits_{{{\left| {h_{m,n}^f} \right|}^2} \le {{\left| {h_{i,n}^f} \right|}^2}\hfill\atop
		\scriptstyle i \in {\mathcal{M}_f}/\left\{ m \right\}\hfill}^{} {p_{i,n}^f}  +\hat I_{m,n}^f+ {\sigma^2} + p_{m,n}^f{{\left| {h_{m,n}^f} \right|}^2}} \right)$ and $g_{m,n}^{f-}=  \log \left( {{{\left| {h_{m,n}^f} \right|}^2}\sum\limits_{{{\left| {h_{m,n}^f} \right|}^2} \le {{\left| {h_{i,n}^f} \right|}^2}\hfill\atop
		\scriptstyle i\in {\mathcal{M}_f}/\left\{ m \right\}\hfill}^{} {p_{i,n}^f}  +\hat I_{m,n}^f+ {\sigma^2}} \right) + \upsilon_{m,n}^{f}.$	
	
We introduce auxiliary variables $t_1$, $t_2$, $t_3$, and $t_4$ to reformulate \eqref{Opt_prob} as  \cite{46}, \cite{47}:

%	\begin{tabular}[b]{|c|c|c|}\hline
%		Scenarios   & Subcarrier allocation    & Power allocatio  
%		\\\hline
%		\multicolumn{1}{|r|}
%		{ASM}
%		&
%		\multicolumn{1}{r|}
%		{	$ O\left( {\frac{{\log \left( {\frac{{F\left( {1 + N\left( {1 + M + M\left( {M - 1} \right) + ME\left( {M - 1} \right) + ME} \right)} \right)}}{{t\partial }}} \right)}}{{\log \left( \xi  \right)}}} \right)$} &\multicolumn{1}{r|}
%		{	$O\left( {\frac{{\log \left( {\frac{{F\left( {1 + N\left( {M + M\left( {M - 1} \right) + ME\left( {M - 1} \right) + ME} \right)} \right)}}{{t\partial }}} \right)}}{{\log \left( \xi  \right)}}} \right)$}
%		\\\hline
%		\multicolumn{1}{|r|}
%		{Monotonic}
%		&
%		\multicolumn{2}{r|}
%		{	$O\left( {F + NF\left( {3M + 1} \right) + \left( {M - 1} \right)\left( {M\left( {M - 2} \right) + 3NFM + 3NFME} \right) + 3FNME} \right)$}
%		\\ \hline
%	\end{tabular}
	 \begin{subequations}\label{monotonic}
		\begin{align}
		&\max_{\mathbf{P},\boldsymbol{\upsilon}, \textbf{T}_1, \textbf{T}_2, \textbf{T}_3}\; \sum_{\forall f\in\mathcal{F}}\sum_{\forall m\in \mathcal{M}_f}\sum_{\forall n\in \mathcal{N}} \left\{ {g_{m,n}^{f+} +T_{4,m,n}^f } \right\},\\&\hspace{.18cm}
		\label{monotonicb}
		 \text{s.t.}:\hspace{.18cm} \hspace{.008cm}
		\eqref{Opt_probb}, \eqref{Opt_probc}, \eqref{Opt_probg},
		\\&
		\label{monotonicc}
		\hspace{.45cm} 
		\hspace{.005cm}{O^ + }\left( {\bf{P}} \right) + T_{1,m,i,n}^f \le {O^ + }\left( {{{\bf{P}}^{\text{mask}}}} \right), 
		 \forall f \in \mathcal{F}, \\& \hspace{.45cm}  n \in \mathcal{N}, m,i \in \mathcal{M}_f, \vert h_{i,n}^{f}\vert^2 \leq \vert h _{m,n}^{f}\vert^2, i \neq m, \nonumber		 
		 \\&\label{monotonicd}
		 \hspace{.45cm} 
		 \hspace{.005cm}
		 {O^ - }\left( {\bf{P}} \right) + T_{1,m,i,n}^f \ge {O^ + }\left( {{{\bf{P}}^{\text{mask}}}} \right)
		 \forall f \in \mathcal{F},	 \\& \hspace{.45cm}  n \in \mathcal{N}, m,i \in \mathcal{M}_f, \vert h_{i,n}^{f}\vert^2 \leq \vert h _{m,n}^{f}\vert^2, i \neq m, \nonumber
			\\ &\label{monotonice}
		 \hspace{.45cm} 
		 \hspace{.005cm}
		 0 \le T_{1,m,i,n}^f \le {O^ + }\left( {{{\bf{P}}^{\text{mask}}}} \right) - {O^ + }\left( {\bf{0}} \right) 
		 \forall f \in \mathcal{F},		 \\& \hspace{.45cm}  n \in \mathcal{N}, m,i \in \mathcal{M}_f, \vert h_{i,n}^{f}\vert^2 \leq \vert h _{m,n}^{f}\vert^2, i \neq m, \nonumber
		\\&\label{monotonicf}
		\hspace{.45cm}
		\hspace{.005cm} 
		{{\hat O}^ + }\left( {\bf{P}} \right) + T_{2,m,i,n,e}^f \le {{\hat O}^ + }\left( {{{\bf{P}}^{\text{mask}}}} \right)
			\forall f \in \mathcal{F}, n \in \mathcal{N}, \\ & \hspace{.45cm}  m,i \in \mathcal{M}_f,    e\in \mathcal{E},  \vert h_{i,n}^{f}\vert^2 \leq \vert h _{e,n}^{f}\vert^2, i \neq m, \nonumber
		\\&\label{monotonicg}
		\hspace{.45cm}
		\hspace{.005cm}
		 {{\hat O}^ - }\left( {\bf{P}} \right) + T_{2,m,i,n,e}^f \ge {{\hat O}^ + }\left( {{{\bf{P}}^{\text{mask}}}} \right)
		\forall f \in \mathcal{F}, n \in \mathcal{N}, \\ 	& \hspace{.45cm}  m,i \in \mathcal{M}_f,    e\in \mathcal{E},  \vert h_{i,n}^{f}\vert^2 \leq \vert h _{e,n}^{f}\vert^2, i \neq m, \nonumber
	    \\&\label{monotonich}
		\hspace{.45cm}
		\hspace{.005cm}
		0 \le T_{2,m,i,n,e}^f \le {{\hat O}^ + }\left( {{{\bf{P}}^{\text{mask}}}} \right) - {{\hat O}^ + }\left( {\bf{0}} \right)
		\forall f \in \mathcal{F}, \\ & \hspace{.45cm} n \in \mathcal{N},  m,i \in \mathcal{M}_f,    e\in \mathcal{E},  \vert h_{i,n}^{f}\vert^2 \leq \vert h _{e,n}^{f}\vert^2, i \neq m, \nonumber
		\\&\label{monotonici}\hspace{.45cm}
		\hspace{.008cm}
		{{\tilde O}^ + }\left( {\bf{P}} \right) + T_{3,m,n,e}^f \le {{\tilde O}^ + }\left( {{{\bf{P}}^{\text{mask}}}} \right)\\&  \hspace{.45cm}  \forall e\in \mathcal{E}, n\in\mathcal{N}, f\in\mathcal{F}, \,\,\, \forall m\in \mathcal{M}_f, \nonumber,
		\\&\label{monotonicj}\hspace{.45cm}
		\hspace{.008cm} {{\tilde O}^ - }\left( {\bf{P}} \right) + T_{3,m,n,e}^f \ge {{\tilde O}^ + }\left( {{{\bf{P}}^{\text{mask}}}} \right)\\&  \hspace{.45cm}  \forall e\in \mathcal{E}, n\in\mathcal{N}, f\in\mathcal{F}, \,\,\, \forall m\in \mathcal{M}_f, \nonumber, 
		 \\&\label{monotonick}\hspace{.45cm}
		 \hspace{.008cm}
		 0 \le T_{3,m,n,e}^f \le {{\tilde O}^ + }\left( {{{\bf{P}}^{\text{mask}}}} \right) - {{\tilde O}^ + }\left( {\bf{0}} \right)		 \\&  \hspace{.45cm}  \forall e\in \mathcal{E}, n\in\mathcal{N}, f\in\mathcal{F}, \,\,\, \forall m\in \mathcal{M}_f, \nonumber,
		 \\&\label{monotonicl}\hspace{.45cm}
		 \hspace{.008cm} T_{4,m,n}^f +g_{m,n}^{f-}\left( {\bf{P}} \right)<=g_{m,n}^{f-} \left({{{\bf{P}}^{\text{mask}}}} \right)
		 \\&\label{monotonicm}\hspace{.45cm}
		 \hspace{.008cm} 0 \le  T_{4,m,n}^f \le g_{m,n}^{f-} \left({{{\bf{P}}^{\text{mask}}}} \right) -g_{m,n}^{f-} \left(\bf{0} \right)
		\end{align}
	\end{subequations}
where, $\bf{P}^\text{mask}$ is a vector which is defined as
${\bf{P}^\text{mask}}
= \left[ p_{m,n}^{f,\text{mask}} \right],
\forall m \in \mathcal{M}_f, n\in \mathcal{N}, f\in \mathcal{F}$, where $p_{m,n}^{f,\text{mask}}$
is the transmit power spectral mask for user $m$ on the $n^{th}$ subcarrier, which is served by the $f^{th}$ BS.
	Moreover, $  {O^ + }\left( {\bf{P}} \right)= p_{m,n}^{f}p_{i,n}^{f} ( |h_{i,n}^f{|^2}{\sigma^2} + {\left| {h_{i,n}^f} \right|^2}{\left| {h_{m,n}^f} \right|^2}\sum\limits_{{{\left| {h_{m,n}^f} \right|}^2} \le {{\left| {h_{l,n}^f} \right|}^2}\hfill\atop
		\scriptstyle l \in {\mathcal{M}_f}/\left\{i  \right\}\hfill}^{} {p_{l,n}^f}+{h_{i,n}^f}\hat I_{m,n}^f)$,  
	$  {O^ - }\left( {\bf{P}} \right)=p_{m,n}^{f}p_{i,n}^{f}( |h_{m,n}^f{|^2}{\sigma^2} +{h_{m,n}^f}\hat I_{i,n}^f+{\left| {h_{m,n}^f} \right|^2}{\left| {h_{i,n}^f} \right|^2}\sum\limits_{{{\left| {h_{i,n}^f} \right|}^2} \le {{\left| {h_{l,n}^f} \right|}^2}\hfill\atop
		\scriptstyle l \in {\mathcal{M}_f}/\left\{ i \right\}\hfill}^{} {p_{l,n}^f})$, 
	  ${\hat O}^+\left( {\bf{P}} \right)= p_{m,n}^{f}p_{i,n}^{f}(|h_{e,n}^f{|^2}{\sigma^2} + {h_{e,n}^f}\hat I_{i,n}^f + {\left| {h_{e,n}^f} \right|^2} {\left| {h_{i,n}^f} \right|^2}\sum\limits_{{{\left| {h_{i,n}^f} \right|}^2} \le {{\left| {h_{l,n}^f} \right|}^2}\hfill\atop
		\scriptstyle l \in {\mathcal{M}_f}/\left\{ i \right\}\hfill}^{} {p_{l,n}^f})$,  ${\hat O}^-\left( {\bf{P}} \right)= p_{m,n}^{f}p_{i,n}^{f}( |h_{i,n}^f{|^2}{\sigma^2}+{h_{i,n}^f}\hat I_{e,n}^f+{\left| {h_{i,n}^f} \right|^2}{\left| {h_{e,n}^f} \right|^2}\sum\limits_{l \in {\mathcal{M}_f}/\left\{ i \right\}}^{} {p_{l,n}^f}, \nonumber)$, ${\tilde O}^ + \left( {\bf{P}} \right)=\log \left( {{{\left| {h_{e,n}^f} \right|}^2}\sum\limits_{i \in {{\cal M}_f}/\left\{ m \right\}}^{} {p_{i,n}^f}  + \hat I_{e,n}^f+{\sigma^2} + p_{m,n}^f{{\left| {h_{e,n}^f} \right|}^2}} \right)$, and ${\tilde O}^ - \left( {\bf{P}} \right)= \log \left( {{{\left| {h_{e,n}^f} \right|}^2}\sum\limits_{i \in {{\cal M}_f}/\left\{ m \right\}}^{} {p_{i,n}^f}  +\hat I_{e,n}^f+ {\sigma^2}} \right)+\upsilon_{m,n}^{f}$.
According to problem   \eqref{monotonic}, we define two sets as follows: 
\begin{align}
&{\aleph _1} = \left\{ {\left( {{\bf{P}},{\boldsymbol{\upsilon }},{{\bf{T}}_1},{{\bf{T}}_2},{{\bf{T}}_3}} \right):{\bf{P}} \preceq {{\bf{P}}^\text{{mask}}}, \eqref{Opt_probb}, \eqref{Opt_probg},} \right.\nonumber \\& \left. {\eqref{monotonicc}, \eqref{monotonicf},  \eqref{monotonici}, \eqref{monotonicl}} \right\},
\end{align}
and
\begin{align}
&{\aleph _2} = \left\{ {\left( {{\bf{P}},{\boldsymbol{\upsilon }},{{\bf{T}}_1},{{\bf{T}}_2},{{\bf{T}}_3}} \right):{\bf{P}} \succeq \textbf{0}, \eqref{Opt_probc}, \eqref{monotonicd}, \eqref{monotonicg}, \eqref{monotonicj},} \right.\nonumber \\&\left. {\eqref{monotonicm}} \right\},
\end{align}
in fact, the
intersection of sets ${\aleph _1}$ and ${\aleph _2}$   is the  feasible set of problem \eqref{monotonic}, moreover, $\aleph _1$ and $\aleph _2$ are normal and co-normal sets, respectively,
in the following  hyper-rectangle, \cite{46}, \cite{47}:
\begin{align}
&\left[ {{\bf{0}},{{\bf{P}}^{{\text{mask}}}}} \right] \times \left[ {0,{O^ + }\left( {{{\bf{P}}^{{\text{mask}}}}} \right) - {O^ + }\left( {\bf{0}} \right)} \right] \times \\& \left[ {0,{{\hat O}^ + }\left( {{{\bf{P}}^{{\text{mask}}}}} \right) - {{\hat O}^ + }\left( {\bf{0}} \right)} \right] \times \left[ {0,{{\tilde O}^ + }\left( {{{\bf{P}}^{{\text{mask}}}}} \right) - {{\tilde O}^ + }\left( {\bf{0}} \right)} \right]\nonumber\\& \times \left[ {0,g_{m,n}^{f - }\left( {{{\bf{P}}^{{\text{mask}}}}} \right) - g_{m,n}^{f - }\left( {\bf{0}} \right)} \right] \times \left[ {{\bf{0}},{{\boldsymbol{\upsilon }}^{\max }}} \right]. \nonumber
\end{align} 
finally,  problem \eqref{monotonic} is a monotonic problem in a canonical
form, based on Definition 5 in \cite{47}. Hence, the optimization problem \eqref{monotonic}
 can be solved by using the polyblock algorithm. As mentioned way, at which we convert \eqref{epi} to  a canonical form of monotonic optimization, we can convert \eqref{p_rho} to a canonical form.
 \subsection{Computational complexity}  
In this section we discuss about the computational complexity of the polyblock algorithm. As we know,  the computational complexity of this algorithm depends on the number of variables and  form of the functions in the optimization problem.
In the  polyplock algorithm  four main steps are performed. In the first step, the best vertex should be found, in the socond step we  find projection of the selected vertex, improper vertexes are  removed in the third step, and new vertex set is found in  the fourth step.
 The dimension of our optimization problem is $
 {\Im _0} = F + NF\left( {3M + 1} \right) + \left( {M - 1} \right)\left( {M\left( {M - 2} \right) + 3NFM + } \right.
 \left. {3NFME} \right)\left.\\ { + 3FNME} \right)$, the convergence of algorithm for stopping threshold $10^{-3}$, occurs  approximately after $\Im_1=10^4 $ iterations, the bisection
algorithm  which gives projection of vertex, with stopping threshold $10^{-3}$,   has   $\Im_2=10^3 $ iterations, approximately. Hence, the complexity order can be written as \cite{moltafet}:
\begin{align}
O\left( {{\Im _1}\left( {{\Im _1} \times {\Im _0} + {\Im _2}} \right)} \right)
\end{align}

\section{SIMULATION RESULTS}\label{SIMULATION RESULTS}

In this section, we provide numerical results to evaluate the performance of the proposed scheme. The simulation parameters are considered as: $p_{0}^{\text{max}} = 16$ dB  (maximum allowable transmit power
of MBS), $p_{m}^{\text{max}} = 6
$ dB, $\forall m \in \mathcal{M}/ \left\{ 1 \right\}$ (maximum allowable transmit power
of SBS), Power Spectral Density (PSD) of noise is $-130$ dBm/Hz,
% $h_{m,n}^f= d_{m,f}^{-\alpha}{\tilde h_{m,n}^f}$, where ${\tilde h_{m,n}^f}$ is the channel fading coefficient,
 $\alpha=4$. Maximum coverage MBS and SBS are supposed 1500 m and 15 m, respectively.

\begin{figure}[h]
		\includegraphics[width=3.2in,height=2.5in]{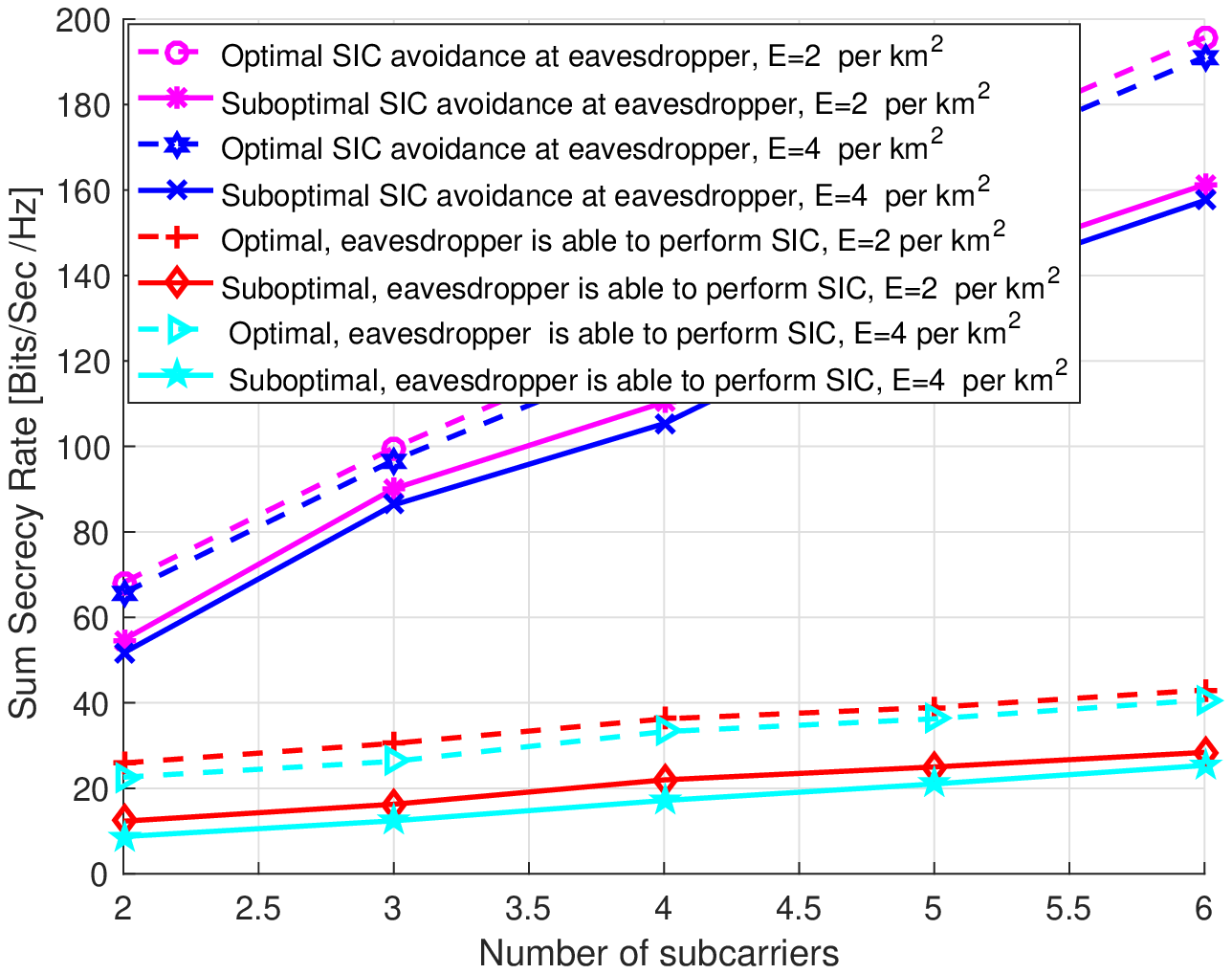}% \vspace{3cm}
		\caption {Secrecy sum rate  versus  the number of subcarriers, comparison between the proposed scheme (SIC avoidance at the eavesdroppers) and when the eavesdropper can do SIC, $M=3$ per km$^2$, $F=2$ per km$^2$.}
		\label{N}
\end{figure}

In Fig. \ref{N}, the sum secrecy  rate versus the number of subcarriers is shown. Also this figure compares our proposed scheme at which the eavesdroppers are not able to perform SIC with the case they can perform SIC. As seen in this figure, the sum secrecy  rate in our proposed scheme has  \%82.52 gap with the conventional type, because in the proposed scheme we do not allow eavesdroppers to perform SIC, even if they know the channel ordering, but in the conventional type the eavesdroppers can perform SIC.  This issue is shown for two different  number of eavesdroppers, i.e., $E=4$, $E=6$.  

\begin{figure}[h]
		\includegraphics[width=3.2in,height=2.5in]{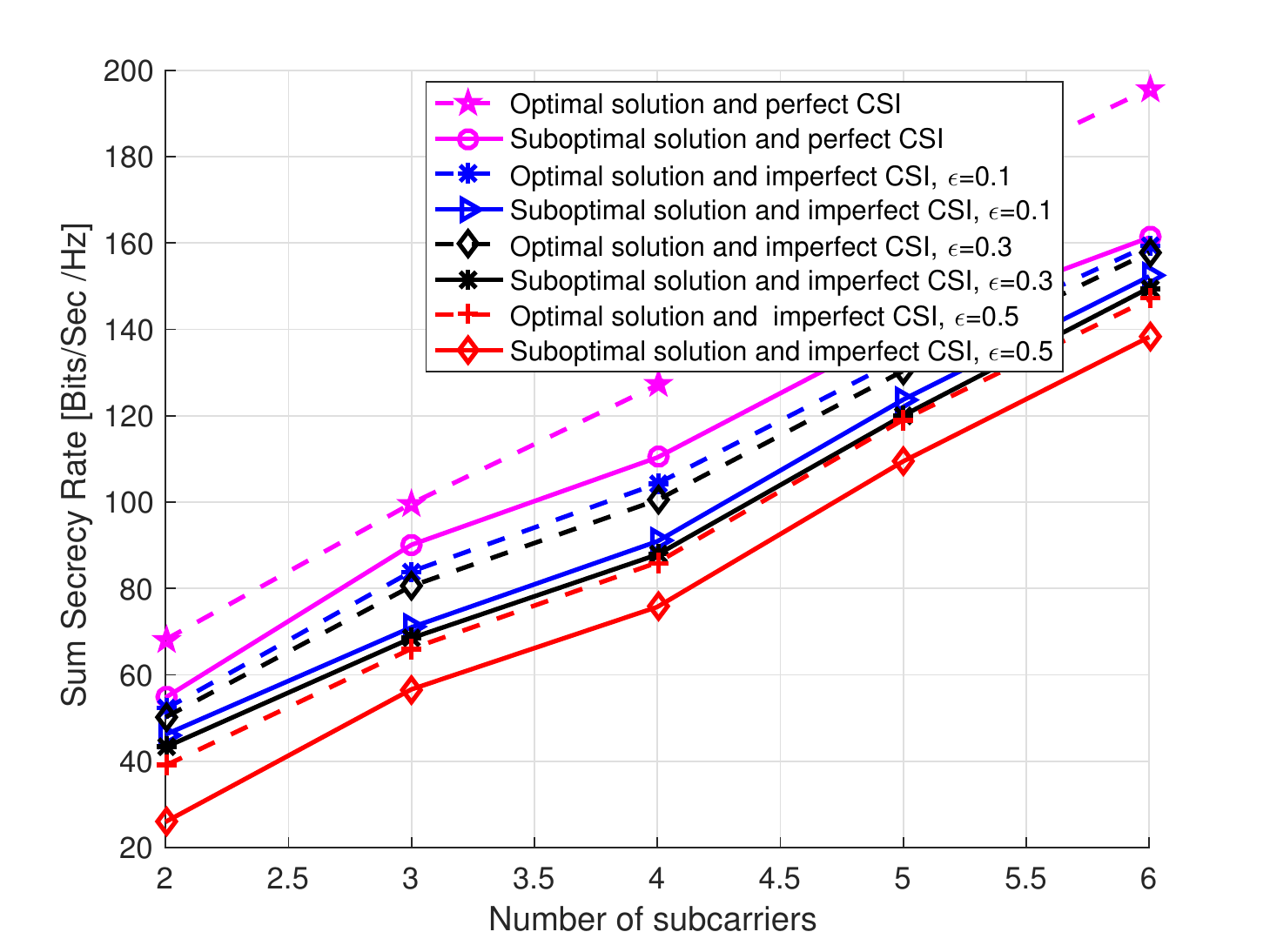}% \vspace{3cm}
		\hspace{-2.8cm}		\caption{Secrecy sum rate  versus  the number of subcarriers for perfect and imperfect CSI in the proposed scheme,  $F=2$ per km$^2$, $M=3$ per km$^2$, $E=2$ per km$^2$.}
		\label{per_imper}
\end{figure}
In Fig. \ref{per_imper}, we compare  the performance of the proposed scheme for the perfect and imperfect CSI scenarios. Moreover, this figure shows imperfect CSI sensitivity with respect to the upper bound of error. As seen, when $\epsilon=0.1$, the imperfect SCI scenario has $\%14.34$ gap with respect to perfect CSI. By increasing $\epsilon$ to $0.3$ and $0.5$, this gap is increased to    $\%17.37$ and  $\%31.45$, respectively.
% As seen, the gap between the knowing the eavesdropper’s CSI and unknowing the eavesdropper’s CSI is increased, when the number of subcarrier increase. 

\begin{figure}[t]
		\includegraphics[width=3.2in,height=2.5in]{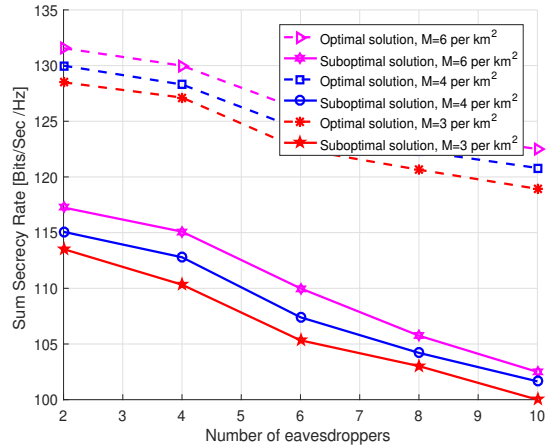}% \vspace{3cm}
		\hspace{-2.8cm}		\caption{Secrecy sum rate versus the number of  eavesdroppers, in the proposed scheme, $F=2$ per km$^2$, $N=4$, $E=2$ per km$^2$.}
		\label{beam}
\end{figure}
Fig. \ref{beam}, shows the sum secrecy  rate versus the number of eavesdroppers. As seen, with increasing the number of  eavesdroppers in our system model, the sum secrecy  rate is decreased. This is because, as we know, it is assumed the eavesdroppers are non-clutsions, therefore, when an eavesdropper is added to the system,  its channel maybe  better than others, hence the secrecy rate changes (decreases). In the simulation, we use the Monte Carlo method, therefore when  the number of eavesdroppers is increased, the secrecy rate decreases on average.

\begin{figure}[h]
		\includegraphics[width=3.2in,height=2.5in]{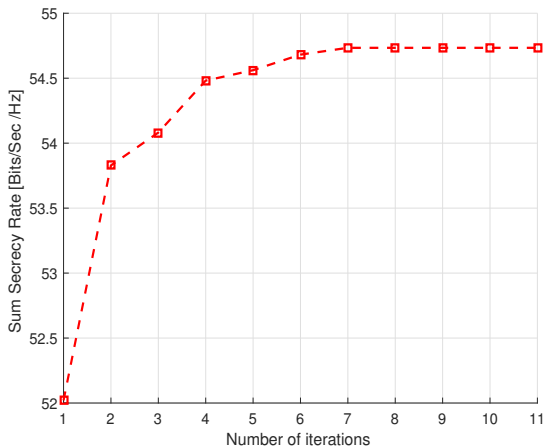}% \vspace{3cm}
		\hspace{-2.8cm}		\caption{Convergence of the proposed algorithm.}
		\label{itr}
\end{figure}
Fig. \ref{itr} presents the convergence behavior of the proposed algorithm. We observe that the algorithm converges in iteration 8, in ASM, approximately. In this figure, we assume $E=2$ per km$^2$, $F=2$ per km$^2$, $M=3$ per km$^2$, and $N=2$.

In addition, we  show optimal solution  in all of these
figures. As seen,  the proposed suboptimal solution which has  low complexity with respect to the  monotonic optimization problem,  
is  closed to the optimal solution, for example in Fig. \ref{beam},  the optimal solution has approximately  \%13.05 gap with the proposed suboptimal solution.

\begin{figure}[h]
	\includegraphics[width=3.2in,height=2.5in]{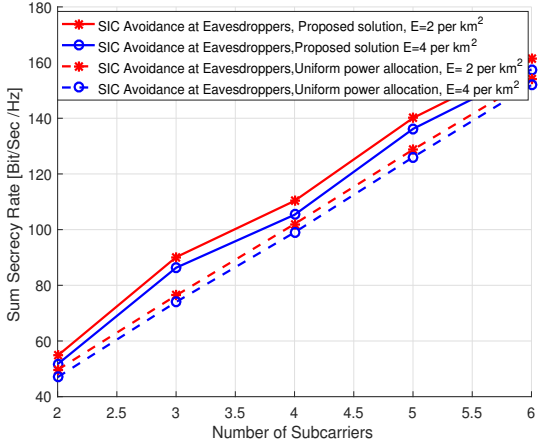}% \vspace{3cm}
	\hspace{-2.8cm}		\caption{Comparison between our proposed solution and the uniform power allocation method, $M=3$ per km$^2$, F=2 per  km$^2$.}
	\label{Comp_P_U}
\end{figure}
\begin{figure}[h]
	\includegraphics[width=3.2in,height=2.5in]{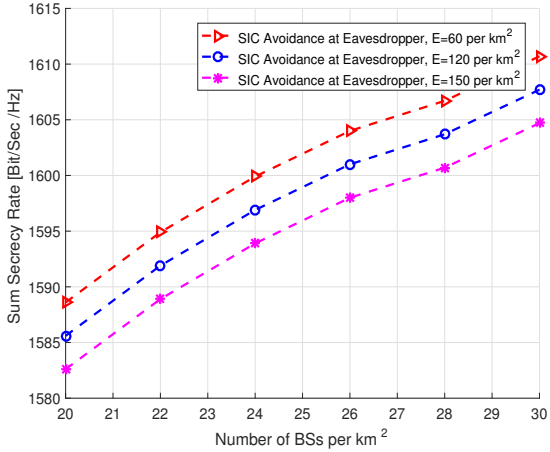}% \vspace{3cm}
	\hspace{-2.8cm}		\caption{Secrecy sum rate versus the number of BSs in HUDN, $N=48$, $M=300$ per km$^2$.}
	\label{HUNDS}
\end{figure}
As mentioned in Section \ref{MCS}, for evaluating the performance degradation due to the uniform power allocation, we compare the uniform power allocation method for a small network dimension with our proposed solution  i.e., joint power and subcarrier allocation in Fig. \ref{Comp_P_U}. As shown in this figure, there is approximately \%9 performance gap between these methods.

In Fig. \ref{HUNDS},  sum secrecy  rate versus the number of BSs in HUDN for massive connectivity is evaluated. Besides, this figure investigates effect of number of eavesdroppers in massive connectivity.
As seen, by increasing one BS per Km$^2$, the  sum secrecy  rate 2.5 unit increases, approximately.

\section{Conclusion}\label{Conclusion}

In this paper, we investigated  physical layer security  for power domain non-orthogonal multiple access based HetNet. We proposed a novel resource allocation   to  maximize the sum secrecy  rate in PD-NOMA based HetNet. In the  proposed scenario, the eavesdroppers  are not allowed to perform successive interference cancellation,  but the legitimate users are able to perform it. Hence,  all users' signals in the eavesdroppers are treated as interference, while some users' signals can be canceled in the desired users, therefore, less interference is experienced by users. In order to solve the optimization problem,  we  adopted the iterative algorithm called ASM, i.e., to convert the optimization problems to two subproblems power and subcarrier allocation and solve them alternatively.  
In each iteration of ASM, we consider fixed subcarrier  and allocated power, then, for subcarrier allocation we consider fixed power.
In each iteration of this method, power allocation problem,
is  non-convex  and subcarrier allocation is NLP. Hence, we  solve the power allocation  by the SCA approach. To this end, we use the DC approximation, to transform the non-convex problem into canonical form of convex optimization. Also, we  solve subcarrier allocation, 
by exploiting MADS, hence we employ the existing solver  called NOMAD.
Moreover, we obtained optimal solution and the optimality gap of the proposed solution method, by converting the optimization problems to a canonical form of the monotonic optimization problem and exploiting the polyblock algorithm. \textcolor{black}{Besides, we evaluated  the proposed scheme for secure massive connectivity in the massive machine-type communication (mMTC) cases in 5G networks. As resource allocation in this networks has high dimension complexity, we allocated  the transmit power uniformly to users and showed the performance of uniform power allocation is close to the performance of our proposed solution.} Numerical results show the sum secrecy  rate in our novel resource allocation  has  \%82.52  gap with the conventional type at which the eavesdroppers are able to perform SIC. Moreover, we investigated imperfect CSI of the eavesdroppers scenario and compared it with the perfect CSI case.
%%%%%%%%%%%%%%%%%%%%%%%%%%%%%%%%%%%%%
%%%%%%%%%%%%%%%%%%%%%%%%%%%%%%%%%%5
\begin{center}
	Appendix I
\end{center}
In the SCA approach with the DC approximation, a sequence of improved feasible solutions is generated and is converged to a local optimum, \cite{local}.

\textit{Proof}: As mentioned, we approximate \eqref{H} with  function \eqref{aprox}. Gradient of function $H_{m,n}^f\left( {\bf{P}} \right)$ is  its super gradient, because $H_{m,n}^f\left( {\bf{P}} \right)$   is a concave function \cite{R9}. 
Hence, we have

\begin{align}
	&H_{m,n}^f\left( {{\bf{P}}\left( \mu  \right)} \right) \le H_{m,n}^f\left( {{\bf{P}}\left( {\mu  - 1} \right)} \right) + \\&{\nabla ^T}H_{m,n}^f\left( {{\bf{P}}\left( {\mu  - 1} \right)} \right)\left( {{\bf{P}}\left( \mu  \right) - {\bf{P}}(\mu  - 1)} \right), \nonumber
\end{align}
Therefore, in the objective function, we have:

\begin{subequations}
	\begin{equation}
	\sum\limits_{\forall f \in {\cal F}} {\sum\limits_{\forall m \in {{\cal M}_f}} {\sum\limits_{\forall n \in {{\cal N}_f}} {\rho _{m,n}^f} } } \left\{ {G_{m,n}^f\left( {{\bf{P}}\left( \mu  \right)} \right) - H_{m,n}^f\left( {{\bf{P}}\left( \mu  \right)} \right)} \right\}\ge \nonumber
	\end{equation}    
	\begin{equation}
	\begin{array}{*{20}{l}}
	{  \sum\limits_{\forall f \in F} {\sum\limits_{\forall m \in {M_f}} {\sum\limits_{\forall n \in {N_f}} {\rho _{m,n}^f} } } \left\{ {G_{m,n}^f\left( {{\bf{P}}\left( \mu  \right)} \right) - H_{m,n}^f\left( {{\bf{P}}\left( {\mu  - 1} \right)} \right)} \right.}\\
	{\left. { - {\nabla ^T}H_{m,n}^f\left( {{\bf{P}}\left( {\mu  - 1} \right)} \right)\left( {{\bf{P}}\left( \mu  \right) - {\bf{P}}(\mu  - 1)} \right)} \right\}}
	\end{array}\nonumber
	\end{equation}
	\begin{align}
	\begin{array}{l}
	= \mathop {\max }\limits_{\bf{P}} \sum\limits_{\forall f \in F} {\sum\limits_{\forall m \in {M_f}} {\sum\limits_{\forall n \in {N_f}} {\rho _{m,n}^f} } } \left\{ {G_{m,n}^f\left( {\bf{P}} \right) - } \right.\\
	\left. {H_{m,n}^f\left( {{\bf{P}}\left( {\mu  - 1} \right)} \right) - {\nabla ^T}H_{m,n}^f\left( {{\bf{P}}\left( {\mu  - 1} \right)} \right)\left( {{\bf{P}} - {\bf{P}}(\mu  - 1)} \right)} \right\}
	\end{array}	\nonumber
	\end{align}
	\begin{align}
	\begin{array}{*{20}{l}}
	\begin{array}{l}
	\ge \sum\limits_{\forall f \in F} {\sum\limits_{\forall m \in {M_f}} {\sum\limits_{\forall n \in {N_f}} {\rho _{m,n}^f} } } \left\{ {G_{m,n}^f\left( {{\bf{P}}(\mu  - 1)} \right) - } \right.\\
	H_{m,n}^f\left( {{\bf{P}}\left( {\mu  - 1} \right)} \right) - {\nabla ^T}H_{m,n}^f\left( {{\bf{P}}\left( {\mu  - 1} \right)} \right)
	\end{array}\\
	{\left. {\left( {{\bf{P}}({\bf{\mu }} - {\bf{1}}) - {\bf{P}}(\mu  - 1)} \right)} \right\}}
	\end{array}\nonumber
	\end{align}
	\begin{align}
&	= \sum\limits_{\forall f \in {\cal F}} {\sum\limits_{\forall m \in {{\cal M}_f}} {\sum\limits_{\forall n \in {{\cal N}_f}} {\rho _{m,n}^f} }} \nonumber \\& \left\{ {G_{m,n}^f\left( {{\bf{P}}\left( {\mu  - 1} \right)} \right) - H_{m,n}^f\left( {{\bf{P}}\left( {\mu  - 1} \right)} \right)} \right\}. \nonumber
	\end{align}
\end{subequations}
Therefore,  after iteration $\mu $, the objective function value either increases or stays unchanged with respect to iteration  $\mu  - 1$.                                                                      

\end{document}